\documentclass[letterpaper,twocolumn,10pt]{article} 
 \usepackage{usenix}

\usepackage{pifont}
\usepackage{pgfplotstable}
\usepackage{paralist}
\usepackage{amsmath}
\usepackage{multirow}
\usepackage{shortcuts}
\usepackage{epsfig,endnotes}
\usepackage{tikz}
\usepackage{pgfplots}
\usepackage{fontawesome}
\usepackage[ruled,vlined,linesnumbered]{algorithm2e}
\usepackage{amsfonts}
\usepackage{url}
\usepackage{tikz}
\usepackage{array}
\usepackage{booktabs,adjustbox}
\usepackage{etex}
\usepackage{color}
\usepackage{pstricks}
\usetikzlibrary{calc,positioning,shapes.geometric,shapes.symbols,shadows,arrows}
\usepackage{wasysym}
\usepackage{shortcuts}
\usepackage{paralist}
\renewcommand{\paragraph}[1]{\vspace{0.1in}\noindent{\bf{#1}.}}
\usepackage[justification=justified]{caption}
\usepackage{pgfplots}
\usetikzlibrary{arrows}
\usetikzlibrary{patterns}
\usepackage{times}
\usepackage{listings}
\usepackage{multicol}
\usepackage{lipsum}
\usepackage{adjustbox}
\usepackage[font=bf,skip=\baselineskip]{caption}
\usepackage{tabularx}
\usepackage{multirow}
\usepackage{shortcuts}
\usepackage{enumerate}
\usepackage[shortlabels]{enumitem}
\usepackage[english]{babel}
\usepackage[utf8x]{inputenc}
\usepackage{listings}
\usepackage{color}
\usepackage{subcaption}
  	
\usepackage{color, colortbl}
\usepackage{hyperref}
\hypersetup{
    colorlinks = true,
    linkcolor = blue,
    anchorcolor = blue,
    citecolor = blue,
    filecolor = blue,
    urlcolor = blue
}

 \usepackage[most]{tcolorbox}
\newtcolorbox[%
auto counter]{lesson}[2][]{%
	enhanced jigsaw,
        colback=pink!20,
	breakable,
	#1}

 \usepackage[most]{tcolorbox}
\newtcolorbox[%
auto counter]{mybox}[2][]{%
	enhanced jigsaw,
        colback=blue!6,
	breakable,
	#1}

\usepackage[all]{nowidow}

 \usepackage{caption}
\begin{document}

% \title{SoK: When Mobile Super Apps Become ``Operating Systems'': \\ The Good, The Bad, and The Ugly} 
\title{SoK: Decoding the Super App Enigma: \\ The Security Mechanisms, Threats, and Trade-offs %The Merits, Pitfalls, and Controversies % Dissecting the Security and Privacy 
%on Apps Morphing into OSes
in OS-alike Apps
}

\author{
{\rm Yuqing Yang}\\
The Ohio State University
\and
{\rm Chao Wang}\\
The Ohio State University
\and
{\rm Yue Zhang}\\
The Ohio State University
\and
{\rm Zhiqiang Lin}\\
The Ohio State University
}

%\title{I Don't Trust Front-ends Anymore:\\ Understanding and Measuring the Misuse of Cryptographic Keys in Mini-Programs}
%\title{Cryptographic Access Control in Mini-Programs:\\ A Closer Look at Key Leakages}
%\author{\#33}
% Unraveling the Benefits, Downsides, and Hidden Complexities

% Problems, Traps, and Pitfalls} % TODO: replace with your title
% Decompiling Java Bytecode: Problems, Traps and Pitfalls

% % TODO: replace this section with code generated by the tool at https://dl.acm.org/ccs.cfm
% <concept_id>10002978.10003029.10011703</concept_id>
% <concept_desc>Security and privacy~Usability in security and privacy</concept_desc>

% \ccsdesc{Security and privacy~Use https://dl.acm.org/ccs.cfm to generate actual concepts section for your paper}
% % -- end of section to replace with generated code

% \keywords{template, formatting, pickling} % TODO: replace with your keywords

\maketitle
\begin{abstract}
%Super app (e.g., WeChat, Baidu) is a novel paradigm where mobile native apps allow third-party developers to deploy add-ons inside them to offer customizable services to end users, resembling the extensions in browsers like Chrome. Due to the versatility and convenience of such a paradigm, the past 6 years has witnessed increasing popularity among more than 20 mobile app giants, such as WeChat in China, Zalo in Vietnam, and Line in Korea. However, now that super app paradigm essentially makes native apps become a Operating System mandating data access between developers and millions of users, it makes these super app and miniapp vendors a favourable target for attackers, despite the profits and convenience the paradigm brings. Over the last three years, researchers have identified various fundamental vulnerabilities in major super app platforms, affecting thousands of miniapps and millions of users. In this paper, we systematically analyze the benefits, drawbacks, and pitfalls of the super app paradigm with the insights and findings we obtained over the past three years, and addressed imminent open problems that need to be addressed. Our results highlight that ....

The super app paradigm, exemplified by platforms such as \textsc{WeChat} and \textsc{AliPay}, has revolutionized the mobile app landscape by enabling third-party developers to deploy add-ons within these apps. These add-ons, known as miniapps, leverage user data hosted by the super app platforms to provide a wide range of services, such as shopping and gaming. With the rise of miniapps, super apps have transformed into ``operating systems'', offering encapsulated APIs to miniapp developers as well as in-app miniapp stores for users to explore and download miniapps. In this paper,  we provide the first systematic study to consolidate the current state of knowledge in this field from the security perspective: the security measures,  threats, and trade-offs of this paradigm. Specifically, we summarize 13 security mechanisms and 10 security threats in super app platforms, followed by a root cause analysis revealing that the security assumptions still may be violated due to issues in underlying systems, implementation of isolation, and vetting. Additionally, we also systematize open problems and trade-offs that need to be addressed by future works to help enhance the security and privacy of this new paradigm.
 % reveal that while various security mechanisms have been enforced at front-end and back-ends, the security assumptions still may be violated and thus introducing security threats, some of which have been confirmed by published documents. Hence, 

%The study conducts a thorough analysis of security measures implemented by super apps, examining both frontend and backend aspects. We noticed that various techniques such as permissions, sandboxing,  encryption, and secure third-party integration are employed to ensure data protection and system integrity. However, potential security threats exist, including phishing attacks, privileged access, cross-platform vulnerabilities, and weak data management. We also highlight the open challenges faced by super apps, such as balancing revenue generation through ads with maintaining a positive user experience, addressing privacy concerns related to fingerprinting attacks, and finding the right equilibrium between usability and security. It is hoped that this study will inspire further research to enhance the security of super apps and miniapps.
 %Achieving user awareness and education is crucial for promoting safe practices within the super app ecosystem.
\end{abstract}

\section{Introduction}
\label{sec:intro}

%\ZY{ read the first few paragraphs of the paper, and I think it discusses extensions too little. Just looking at the title, I expected the paper to first explain how mini-apps evolved from web extensions.}

% \AY{The comments Dr Lin gave last time:}
% \begin{itemize}
%     \item Motivate well this point:
%     \begin{itemize}
%         \item People can live inside super app entirely without going to other apps
%         \item platform can earn money from the \textbf{ecosystem}
%     \end{itemize}
%     \item Define what is OS, what is resource, and a table of what kind of systems manages what kind of resources
%     \begin{itemize}
%         \item DOS: local resources
%         \item Cloud system: remote resources, resources on the servers
%         \item Android/iOS: Local resources (Android), cloud-stored user resources (apps)
%         \item Miniapp: Local resources (super app), cloud resources (super app and miniapp)
%     \end{itemize}
% \end{itemize}
% \AY{Comments section end}

%Miniapp~\cite{MiniAppS51:online} is an innovative paradigm that emerged from the convergence of web and mobile principles. It revolves around a mobile native application, referred to as the super app, which empowers third-party developers to create add-ons for the benefit of super app users. 

%\AY{in introduction, raise the key insight clearly, followed by which show research questions }
The super app paradigm pioneered by \textsc{WeChat}
% , exemplified by platforms such as WeChat and Baidu,
has emerged as a convenient way for mobile applications catering to massive amount of users. It enables seamless integration and management of third-party services, which are developed and submitted by external developers as add-ons. These third-party services typically take the form of miniapps (self-contained code packages submitted to super apps) that are distributed to and executed within end users' host apps. By leveraging native services like payment and accessing user data stored within the super app, these miniapps greatly expand the range of services inside super apps, effectively transforming the super app into an app ecosystem akin to an operating system. 

As a versatile solution for super apps to provide customizable and light-weight services to end users, an increasing number of mobile giants are transforming into super apps in seek of a novel channel for increasing users' stickiness and also monetary profits. For instance, \textsc{WeChat} (a super app with 1.2 billion users) debuted the ``miniapp'' paradigm in 2017. In Tencent's Q3 2023 report, \textsc{WeChat} have hosted more than 3.5 million miniapps, providing services to over 600 million Daily Active Users (DAU), and generating over 2.7 trillion RMB in transactions~\cite{wechatreport2023}. Witnessed by the profits generated from this novel paradigm, there had been more than 20 top mobile app platforms that transformed into super apps% As a new way for giant mobile platforms to build ecosystems and generate additional profits, there had been more platforms 
% Starting from 2017, there have been more than 20 top major super apps
, and this includes \textsc{WeChat} in China (social, 2017)~\cite{wechat3M}, \textsc{Kakao} in Korea (social, 2017), \textsc{Rakuten} in Japan (shopping, 2019), and \textsc{Zalo} in Vietnam (social, 2021)~\cite{superapps2021mau}, since this paradigm has made these apps ``\textit{sort of like Twitter, plus PayPal, plus a whole bunch of other things. And all rolled into one great interface}'', as mentioned by Elon Musk~\cite{musksuperapps}. \looseness=-1

With the advent of miniapps, super apps have transformed into comprehensive ``operating systems'' 
% in their own right. In this paradigm, the super app 
by integrating an execution environment along with encapsulated APIs, such as \texttt{getUserInfo}~\cite{getuserinfo}, which manages resource access from miniapps to user account information from the super app. These miniapps are essentially web front-ends that are packaged to run on customized execution engines provided by the super app platforms. They leverage super-app-specific JavaScript and Cloud APIs to deliver their functionalities. This parallels traditional mobile  operating systems like Android or iOS, where the OS providers offer the environment and programmable APIs. Additionally, there is an in-app miniapp store, akin to platforms such as \textit{Apple AppStore} and \textit{Google Play}, where users can explore and download a diverse array of uploaded and vetted miniapps.

%Technically, the super app resembles  ``operating systems'' in three aspects. First, the super apps integrate webview environment for miniapps (which are essentially packed web apps) to be executed within, which essentially makes the super app a ``sandbox'' for third-party code execution. Second, the super apps provide and manage resource access with various encapsulated APIs, including those managed locally on users' devices like Bluetooth\AY{cite bluetooth}, and those managed on super app cloud, such as user's account information \texttt{getUserInfo}~\cite{getuserinfo}. Third, the super apps maintain in-app miniapp store akin to platforms such as \textit{App Store} and \textit{Google Play}, where the platforms control and vet the distribution of miniapps against malware, and users can freely explore and download miniapps they need.

However, now that the third-party has been given the access to super app managed resources, including a vast amount of user information, cloud services, and OS resources, it is crucial to understand the security challenges and risks it posts to the super app platform and its end users, as the security threats exploited by attackers may affect massive amount of users. To this end, in this paper, we present the first systematic study to reveal the security and privacy of super app paradigms by thoroughly  scrutinizing and dissecting the protection mechanisms, security threats, and their root causes. To be more specific: 
% However, 
% Now that the miniapp paradigm has firmly established itself as an operating system, it is crucial to conduct comprehensive research that analyzes its benefits, addresses its negative aspects, and particularly focuses on the security challenges it presents. The reason for this urgency lies in the vast amount of sensitive resources, such as user information, OS resources, and super apps' cloud services, that are collected and utilized by miniapps. It is imperative to protect these resources from any form of abuse or improper manipulation. 
% to the best of our knowledge, no prior works have  examined and summarized the security and privacy aspects of this paradigm. Therefore, there is a pressing need for new research efforts to thoroughly understand, scrutinize, and dissect the miniapp paradigm. In response to this call, we propose the first systematic study that aims to consolidate the current state of knowledge in this field:

\begin{packeditemize}
 \item \textbf{Evolution and Taxonomy (\S\ref{sec:back}).} We analyze the historical evolution of the super app ecosystem and miniapps, comparing them with web apps and native apps to highlight the recent advancements and difference from traditional paradigms. Our study provides the first taxonomy of super apps based on the architecture and implementation of environments that support the execution of miniapps. 
 \item \textbf{Security Mechanisms (\S\ref{sec:good}).} We conduct a comprehensive analysis of 13 security measures enforced by super apps through the lens of the communication between miniapp front-ends, back-ends, and super apps.
 % examining both front-end and back-end aspects, including the local resource management, miniapp execution protection, and cloud communication protection mechanisms.
 These mechanisms include those adopted from traditional paradigms and adapted to super apps (e.g., sandboxing), enforced to isolate access ot services and data from unexpected parties (e.g., API restriction), and implemented to scrutinize third-party code for thwarting malware (e.g., code vetting). \looseness=-1 
 % At the frontend, super apps utilize strong permissions, sandboxing, Document Object Model (DOM) tree isolation, hot update restrictions,   and API restrictions to ensure data protection and system integrity.  At the backend, they employ domain allowlisting, secure communication, token-based services access, data encryption, RBAC~\CW{Role-based access control?}, code vetting, account protection, and secure third-party integration to guarantee trusted content access, encrypted transmission, controlled resource access, secure data handling, controlled user privileges, code verification, account safeguarding, and secure integration within the app ecosystem.

\item \textbf{Security Threats (\S\ref{sec:bad}).} We systematically analyze the security threats by summarizing the security assumption on each security mechanisms and potential risk of violation. Consequently, we identify 10 threats that have been confirmed by published documents including research papers, security reports, and announcement from super app platforms. The root causes to these attacks include compatibility issue caused by different implementation of security mechanisms in underlying OSs, implementation issue in super app security mechanisms, trust issue between the platform and developers, and vetting issue due to limitation of the vetting mechanisms.
  \item \textbf{Lessons Learned (\S\ref{sec:lessons}).} Based on the security threats and root causes, we systematically compile our observations into 5 lessons learned on the challenges super apps face that may provide insight for designing security mechanisms and mitigating the security threats. 
  \item \textbf{Open Problems (\S\ref{sec:controversies}).} Based on the security analysis and lessons we summarized, we identified trade-offs super app platforms face and need to address, as well as 4 open problems, including performing automated security analysis, standardizing security mechanisms, educating developers on security configurations, and developing semantic-aware miniapp vetting. These open problems may be addressed by future works to help enhancing the security and privacy of super app platforms.
\end{packeditemize}

 \section{Super Apps in a Nutshell}
\label{sec:back}
 
 \subsection{Characteristics of Super Apps}% Communication Models}
%\AY{ briefly discuss how miniapps evolved from web apps and in-app browsers. On top of that, show we collected how many super apps and how we categorize them. Also, show the figure of different architectures}

% A super app is a single mobile application that combines multiple services and features into one platform, providing users with a wide range of functions. It eliminates the need for separate apps for tasks like messaging, ride-hailing, food delivery, shopping, and banking. Super apps utilize miniapps, which are add-ons that enhance the user experience and combine the benefits of web and native apps. As shown in \autoref{fig:paradigm},  in the super app paradigm, the app operates on top of the device's operating system, managing both local and cloud resources. It facilitates communication with miniapps developed by third-party developers using provided tools and APIs. The super app grants access to sensitive user data, with permission prompts to ensure user control. Miniapp developers undergo vetting processes and publish their miniapps on the super app platform's marketplace to maintain ecosystem security.

A super app is a mobile application that integrates multiple services offered by third-parties, providing users with a diverse array of functions within a unified app. Unlike traditional approaches that require separate applications for distinct purposes like messaging, ride-hailing, and shopping, a super app integrates these services into a centralized platform by allowing third-party developers to integrate miniapps inside the super app. These services can be accessed by users on-demand via built-in miniapp  store, without requiring to install the miniapps on users' mobile devices. In general, these miniapps are akin to Progressive Web Applications (PWA) as derived from the W3C standards, which include HTML, CSS, JavaScript (JS) scripts, as well as manifest files to configure functional and security mechanisms. However, encapsulated APIs are provided by super apps for miniapps to access a variety of native services and user data managed by the super apps. Hence, compared with web apps, the miniapps have more powerful capabilities to interfere with System- and Superapp-specific resources.
% because the JS SDKs, markup languages, and style sheets are cutsomized by the super apps to support native access.
For example, a miniapp can directly launch in-app payment or pair Bluetooth devices via miniapp JS API, which is handled by the super apps through an interface layer called JavaScript bridge. \looseness=-1

\begin{figure}
    \centering
    \includegraphics[width=\linewidth]{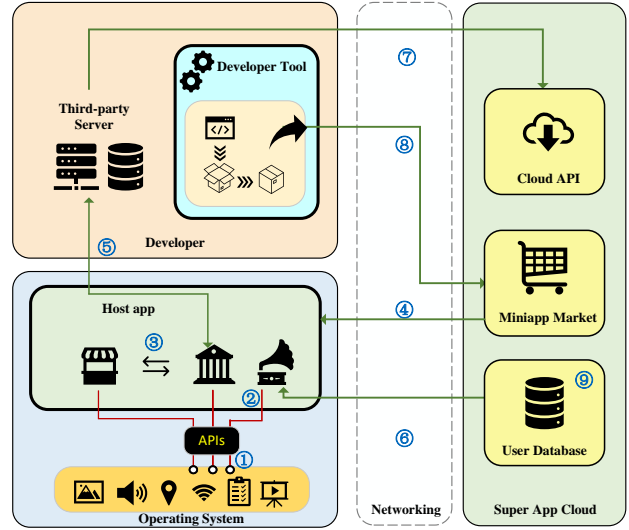}
    \caption{Overview of the key components of super apps}
    \label{fig:paradigm}
\end{figure} 
% Powered by these access to resources
As third-party developers are involved in this paradigm, the communication of the super app paradigm 
% With the introduction of third-party developers, the super app's communication 
generally involves three parties as depicted in \autoref{fig:paradigm}: the developers (orange box), the super app cloud (green box), and the host apps that are downloaded to mobile devices of end users (blue box). On end devices,
% As depicted in \autoref{fig:paradigm}, 
host apps operate on top of the underlying operating system on end users' devices, managing both local OS resources available to the super apps (such as Bluetooth and GPS location) and cloud resources hosted by super apps (such as the phone numbers and addresses of users). Access to both local and cloud resources is then made convenient for developers, as they only need to invoke these highly-encapsulated APIs without having to worry about underlying mechanisms.
% With these resources, the super apps provide APIs for developers to access these local and cloud resources, enabling convenient implementation of versatile services for developers utilizing the developer tools and miniapp SDK provided by super apps, without requiring these developers to worry about underlying mechanisms. 
For instance, when accessing location data, developers only need to invoke \texttt{getLocation()}, without having to implement mechanisms to prompt the users asking for permission, as these are performed by super apps without interference from developers.
% , which are developed by third-party developers utilizing he provided development tools and programmable APIs. These APIs can encompass both local APIs for accessing local resources and cloud APIs for accessing cloud resources. Under the supervision of the super app, miniapps gain access to sensitive user data. For instance, when attempting to access location data, the super app prompts the user for permission, empowering them to make informed decisions. 
However, to ensure the security of the ecosystem, registering developer accounts and providing proof of identity is required. These developers are being rigorously vetted to prevent malicious developers from sabotaging the platform. \looseness=-1
% miniapp developers are required to register accounts, undergo rigorous vetting processes imposed by the super app platform to prevent the uploading of malicious software, and ultimately publish their miniapps on the platform's marketplace.

\subsection{Taxonomy and Evolution}

\begin{figure*}[ht]
    \centering
    \includegraphics[width=0.9\linewidth]{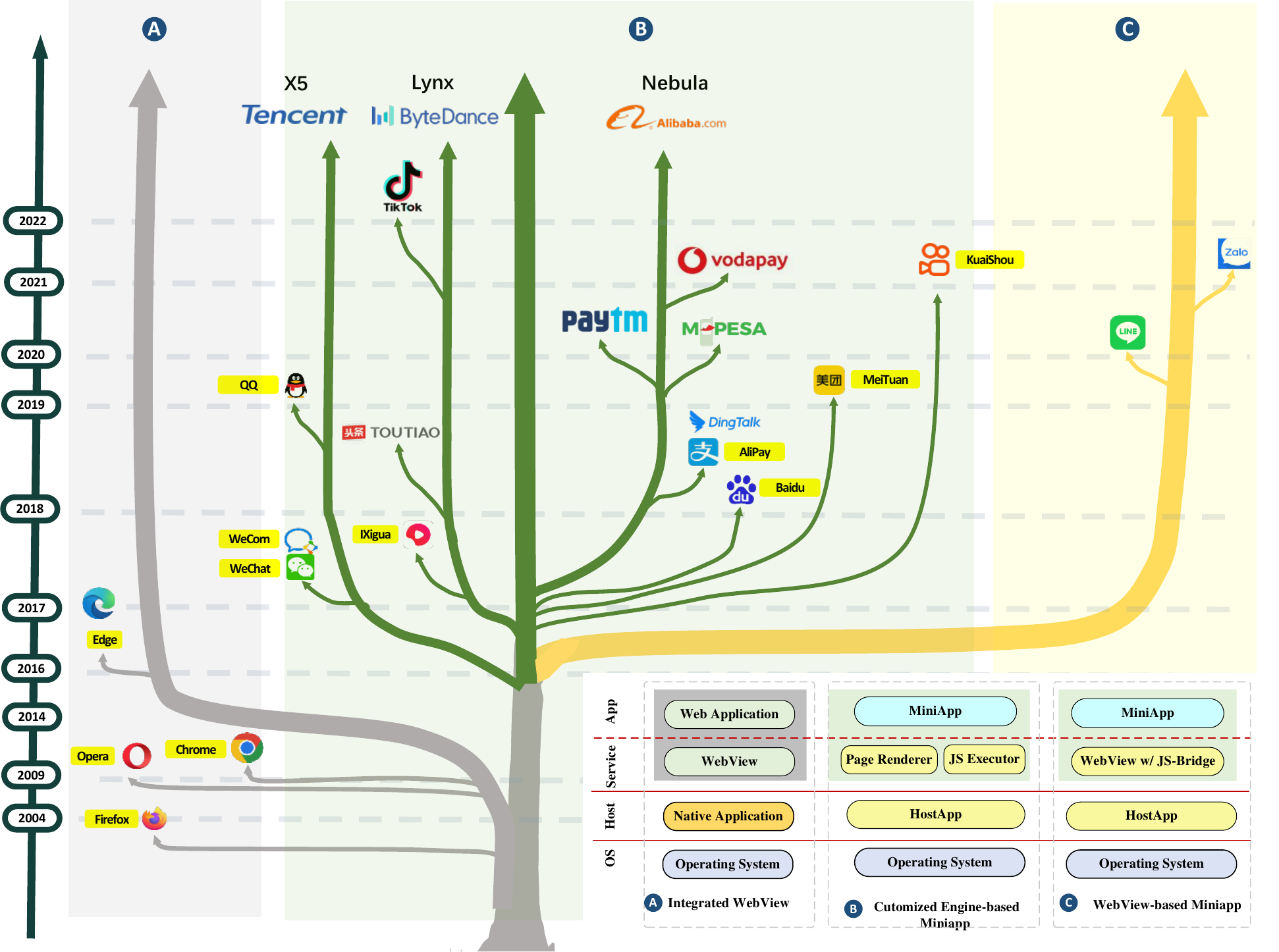}
    \caption{The taxonomy of super app platforms. The architectures of each cluster are shown at the bottom right, where resources in  boxes with dotted lines are optional.}
    \label{fig:evolution}
\end{figure*}

While \textsc{WeChat} introduced the concept of miniapps in 2017, similar paradigms of integrating add-ons into standalone apps have existed prior to that. For example, Google Chrome introduced web extensions in 2009~\cite{chromeextension09}, allowing developers to submit add-ons for the browser. Pioneered by these browsers, when super apps seeks to enable third-party service integration into a single app and to address the fundamental problem of providing cross-platform execution support, the W3C standard have become a feasible solution to follow, because the techniques adopted by web apps and browsers from W3C standards have been developed and tested over the years of web browsing, and these standardized solutions support web app browsing across platforms by design, which suits the needs of super apps. Hence, these super apps followed the techniques used by W3C such as JavaScript, HTML, and CSS, although certain customization have been made to allow native capabilities.   In the past five years, the concept of a super app, i.e., native apps supporting integrated add-ons, has extended beyond \textsc{WeChat}. As shown in \autoref{fig:evolution}, this paradigm has been adopted by over 20 platforms providing diverse services. Among these super app platforms, the foremost notable feature, as well as the top-level classification criteria, is how the execution environment is implemented to support the add-ons or miniapps to free the miniapp developers from concerns about compatibility across the underlying operating systems (Android, iOS, or Windows).
% its multi-platform compatibility, eliminating concerns about the underlying operating system (Android, iOS, or PC). 
As shown in \autoref{fig:evolution}, there are three clusters of solutions adopted by super apps: browser-native environment, cross-platform-framework environment, and super-app integrated environment. \looseness=-1
\begin{packeditemize}
    \item \paragraph{(A) Integrated WebView} Browsers such as \textit{Chrome} and \textit{Opera} began providing APIs for extensions to access services provided by their browser kernels in 2009. These extensions are essentially packages distributed to end users' devices containing JavaScript, HTML, and CSS files, along with a manifest file specifying information about the extension and permission required, which can interact with users' browsers for more powerful functionalities compared with normal web apps. For instance, upon launching an extension, the injected \texttt{chrome} object provides developers with access to functionalities such as monitoring the active tab being browsed by users. Meanwhile, the super apps extend their services by integrating browser kernels to support in-app web browsing, with customized JavaScript objects to extend the functionalities as browsers do, marking the precedent of the miniapp paradigm. \looseness=-1

    \item \paragraph{(B) Customized Engine-based Miniapps} While in-app browsing enable users to visit web pages providing various services without having to open an external browser app, the performance and security have long been an issue for these super apps. First, opening web pages require the super apps to load resources each time, resulting in extended loading time showing blank pages to end users, especially when the network is slow or unstable, significantly undermining user experiences~\cite{wechatarch1}. Meanwhile, the super apps could not vet the third-party websites for malicious code, which may allow attackers to exploit the super apps. To address these problems, super apps including \textsc{WeChat} began to optimize the execution environment by separating logic execution and rendering into different threads to optimize the multi-procedural capability of mobile devices. Meanwhile, third-party developers are forced to submit packed web apps to the WeChat platform, and these packages are only released after being vetted, which significantly reduces the threats from malware. As illustrated on the green arrows in \autoref{fig:evolution}, while different companies implement different versions of execution engines, the separated and customized execution environment implemented by these super apps remain identical, and some company's frameworks (Nebula from Alibaba) are even adopted in super apps in Africa (MPesa and Voda Pay).
    \item \paragraph{(C) Webview-based Miniapp} Apart from platforms that customize and refactor the entire execution environment, these super apps including \textsc{Zalo} directly adopt Android WebView to support the execution, while still requiring developers to submit miniapps to the platform to be vetted. Instead of letting users download the packages, these packages are hosted online on the platform's CDN (e.g., \url{h5.zdn.vn} for \textsc{Zalo}). However, while these packages are not distributed to end users' devices as type (B) super apps do, the miniapps are still vetted by the platforms, as the developers still have to submit their packages to the platforms before these packages are hosted on CDNs of super apps. Hence, whether third-party code is vetted and executed within the application is a core criterion to differentiate traditional paradigms from super app paradigms.

\end{packeditemize}

% \begin{figure*}
%     \centering
%     % \includegraphics[width=\linewidth]{image/1111.png}
%     \includegraphics[width=\linewidth]{image/croppedarch.pdf}
%     % \caption{Replace this with the architectures of different types of super apps\AY{No. This one is still less interesting compared with what the upper one shows}}
%     \label{fig:comparison}
% \end{figure*}

% As illustrated in \autoref{fig:comparison}, in this paper, extension-based super apps refer to the super apps who support native execution and rendering of web pages, and apply them to packed `websites', which become extensions. Plugin-based super apps like Atom who utilize cross-platform frameworks such as Electron or NW.js to enable plugins to be loaded in the super app. Miniapp-based super apps like WeChat refer to the super apps who are not initially designed for running webapp-like add-ons, but later supported the execution by integrating a customized browser kernel into the super apps.
% \paragraph{(2) } 
\ignore{
Interestingly, WeChat, as an example of a super app, has undergone various transformations in different stages. Initially (2011-2014), it operated as a Browser-based super app, allowing users to open in-app web pages and enabling third-party websites to extend functionalities by invoking native APIs through the \texttt{WeixinJSBridge} integration. In the subsequent stage (2015-2016), it evolved into a Cross-platform-framework-based super app with the introduction of a JS-SDK, simplifying development without in-depth knowledge of the underlying implementation. However, challenges of performance and security persisted. To address these concerns, the paradigm shifted towards miniapps (2017-2023), where WeChat became an Integrated-JS-engine-based super app. Miniapps leverage native APIs and resources within the super app ecosystem for optimized performance and faster responses, focusing on specific functionalities to enhance overall performance. Throughout these stages, WeChat underwent transformations to provide a seamless user experience while addressing performance and security concerns.
}

\subsection{Comparision}
% Like webpages and PWAs, miniapps are also derived from W3c standards, but the miniapps are different.

% Distribution: packed, app store

% Execution: customized engine extended W3C standard, often split rendering and execution. Customizations more focused on native/super-app resources access and security requirements. Based on rendering machines and implementations, the miniapps may be running directly in OS (native webview), super app (customized webview), or other frameworks (e.g., Nebula)

While sharing similarities with W3C-derived paradigms such as web pages and PWAs, super apps present unique characteristics and serve specific purposes from the perspectives of platforms, developers, and end users: \looseness=-1

\paragraph{Platforms} As the critical component to support third-party code execution, super apps have unique differences compared with traditional systems. As described in \autoref{tab:cmp3apps}, we  use Android, Chrome, and WeChat as examples to illustrate the core differences between these platforms.
\begin{packeditemize}
    \item \paragraph{Code Distribution} In web pages, the codes (including HTML, CSS, and JavaScript) are distributed via network and referred to online, whereas in native operating systems, app codes are generally distributed as packages, which can be downloaded and installed. Similar to native systems, super apps adopt the packed and self-contained means of app distribution. % to ameliorate the experience for users in weak connection environments (because users do not need to download the packages again the second time they use a miniapp). 
    However, compared with Android system, super apps seek more strict control on the distribution of miniapps to prevent malware distribution. To achieve this, super apps prohibit the users from installing miniapps from third-party sources, and all miniapps must be accessed via the exclusive in-app app store. \looseness=-1
    \item \paragraph{Managed Resources} Traditional Operating Systems like Android mainly manage local resources on the users' devices such as Bluetooth connection and storage. However, super apps are built upon these resources and also allow developers to access the cloud-stored user data. Meanwhile, the super apps feature exclusive app store, which provides the super apps means to manage and vet miniapp packages uploaded by third-party developers. Hence, these packages also automatically become managed resource of super apps. On top of that, a fundamental difference between super apps and other paradigm is that registering an account is mandatory for using or developing miniapps in super apps, which grants super apps the ability to manage these accounts as well.\textit{ In short, the cloud-stored data, miniapp packages, and super app services differentiate super apps from traditional systems in terms of additional types of managed resources.}
    
    % As such, super apps distinct themselves from traditional systems for the additional responsibility and capability to manage these cloud resources from malicious parties.
    \item \paragraph{Execution Environment} While the miniapps running in super apps are derived from the similar technologies from W3C standards, such as JS, HTML, and CSS, the miniapp execution environments are heavily customized and extended beyond the W3C standards by the super apps, with a highlight on access to super app and local resources as mentioned in managed resources. Consequently, such customization makes the execution environment to grant miniapps more native capabilities to interfere with sensitive data and resources.
\end{packeditemize}

% And they have some fundamental differences.,

% Super apps thrive by incorporating miniapps, mobile apps, and web apps, each exhibiting distinctive characteristics and purposes.  In terms of development, there are multiple differences: First, miniapps and web apps are crafted using web technologies such as JavaScript, HTML, and CSS, whereas mobile apps make use of platform-specific languages and frameworks. Second, miniapps and, particularly, web apps have restricted APIs and functionalities, contingent upon web technologies and browser capabilities. In contrast, mobile apps provide an extensive array of features, capitalizing on the hardware and software of the device. Third, miniapps seamlessly integrate within the designated super app or platform, while mobile apps are tailored for specific mobile operating systems. On the other hand, web apps are independent of platforms and can be accessed through web browsers on various devices. Lastly, prior to being made available to the public, miniapps must undergo evaluation by the super app providers to ensure that end users are shielded from any potentially harmful or malicious miniapps in the market.
 
\paragraph{Developers} As the underlying execution engines are extended web execution engine, from the developer's perspective, developing a miniapp is similar to web apps, but with a more customized set of JS APIs and UI components. Contrary to web apps whose front-end components and scripts have to be delivered from a web domain to users' browsers, the developers of miniapps do not necessarily need to maintain a back-end for delivering the front-end code, as the code is submitted and hosted by the super apps. Moreover, even if some developers require cloud back-ends to process data, the super apps still provide cloud database and cloud function features for developers, serving as a means for super apps to generating more profit via cloud services. Hence, developers do not need to have a back-end to develop a miniapp. \looseness=-1

\paragraph{End Users} From an end user's perspective, miniapps, mobile apps, and web apps offer different experiences and functionalities. Miniapps are typically dependent on a super app for updates, and their installation occurs within the super app itself. Mobile apps, conversely, rely on app stores for updates and are installed as individual entities on devices. 
In contrast, web apps update seamlessly via web servers and are directly accessible through web browsers. Performance-wise, the efficiency of miniapps varies according to the underlying super app or platform. Mobile apps, however, can capitalize on device-specific hardware optimization, providing high-performance experiences. The performance of web apps hinges on factors like Internet connectivity, browser capabilities, and server response times. Despite these variables, native apps usually offer superior speed and responsiveness.

In terms of offline functionality, miniapps may have limited capabilities based on the super app or platform, whereas mobile apps can offer extensive offline features. Traditional web apps typically necessitate an Internet connection for functionality, but progressive web apps (PWAs) provide some offline functionality through caching and service workers. 
Regarding user login and registration, users of browsers and Android do not have to register account prior to accessing extensions and apps, whereas super apps cannot be used without registering an account, as accounts are considered a resource hosted by super apps, and super apps enforce account management against malicious users.
% this is a common feature in both mobile and web apps, facilitating personalized experiences and secure access to specific features or content. Miniapps, on the other hand, don't always require this, as they can leverage user information from the super apps. 

\subsection{Threats to Validity} It is worth to note that the super app paradigm is a novel paradigm and super app platforms are still emerging rapidly across the globe. Although certain platforms such as \textit{Grab} label themselves as super apps, their extension of services is mostly confined to implementing individual webviews on their own. They have yet to fully embrace third-party developer integration within their apps. \textit{Therefore, our paper primarily focuses on those super app platforms that not only enable third-party development but also exclusively host and vet the packages. These characteristics distinguish them from traditional paradigms, making them core to our investigation.} To this end, we have identified 15 super apps that fit these criteria, and a more comprehensive list of the platforms we've examined can be found in \autoref{tab:list} of \autoref{sec:yeye}. Also, to illustrate the security measures and potential threats associated with super apps, we have chosen to use {\sc WeChat} as an example in the rest of this paper, given its widespread popularity.

\begin{table}[t]
\centering
\scriptsize
\scalebox{1}{
\setlength\tabcolsep{1pt}
\begin{tabular}{@{}lccc@{}}
\toprule
 
              \textbf{Hosts}            & \multicolumn{1}{c}{\textbf{\begin{tabular}[c]{@{}c@{}} Mobile OS\\(Native Apps)~~~\end{tabular}}}  & \multicolumn{1}{c}{\textbf{\begin{tabular}[c]{@{}c@{}}  Web Browsers~~~\\(Web Apps)~~~ \end{tabular}}} & \multicolumn{1}{c}{\textbf{\begin{tabular}[c]{@{}c@{}} Super Apps\\(Miniapps)\end{tabular}}} \\ \midrule
              ~\textbf{Example Platform}&Android&Chrome&WeChat\\\midrule[1.2pt]

\multicolumn{4}{l}
{\textbf{Platforms}}                       \\ \midrule
~~~\begin{tabular}[]{@{}l@{}}\textbf{Managed resources}\end{tabular}    \\ 
~~~~~~~\begin{tabular}[]{@{}l@{}}Local Resources\end{tabular}       \\
~~~~~~~~~~~\begin{tabular}[]{@{}l@{}}System Resources?\end{tabular}     &   \faCircle  & \faAdjust     &  \faCircle      \\ 
~~~~~~~~~~~\begin{tabular}[]{@{}l@{}}Super-app Services?\end{tabular}          &   \faCircleO  & \faAdjust          &  \faCircle      \\ 
~~~~~~~\begin{tabular}[]{@{}l@{}}Cloud Resources\end{tabular}  \\  
~~~~~~~~~~~\begin{tabular}[]{@{}l@{}}User Data/States?\end{tabular}          &  \faAdjust & \faCircle          & \faCircle        \\  
~~~~~~~~~~~\begin{tabular}[]{@{}l@{}}Account?\end{tabular}          &   \faCircle  & \faCircle          &  \faCircle      \\  
~~~~~~~~~~~\begin{tabular}[]{@{}l@{}}App Packages?\end{tabular}          &   \faCircleO  & \faCircleO           & \faCircle        \\ 
~~~~~~~~~~~\begin{tabular}[]{@{}l@{}}Cloud Services?\end{tabular}          &  \faAdjust & \faCircle          & \faCircle        \\  
~~~\begin{tabular}[]{@{}l@{}}\textbf{App Distribution}\end{tabular}    \\ 
~~~~~~~\begin{tabular}[]{@{}l@{}}App Store?\end{tabular}          &   \faCircle  & \faCircleO     &  \faCircle      \\ 
~~~~~~~\begin{tabular}[]{@{}l@{}}Exclusive?\end{tabular}          &   \faAdjust  & \faCircleO         &  \faCircle      \\ 
~~~~~~~\begin{tabular}[]{@{}l@{}}Packed?\end{tabular}          &   \faCircle  & \faCircleO     &  \faCircle      \\ 
~~~\begin{tabular}[]{@{}l@{}} \textbf{Environment}\end{tabular} & Android Kernel & Web Engine & \begin{tabular}[]{@{}l@{}}Web Engine \\(customized)\end{tabular}  \\ 
\midrule[1.2pt]

\multicolumn{4}{l}
{\textbf{Developers}}                       \\ \midrule
~~~\begin{tabular}[]{@{}l@{}}Language?\end{tabular}          &  C,Java, XML   & JS, HTML, CSS         & JS, WXML, WXSS        \\  
~~~\begin{tabular}[]{@{}l@{}}API Support?\end{tabular}          &  Rich   & Poor         &   Median      \\  
~~~\begin{tabular}[]{@{}l@{}} Compatible with Platforms?\end{tabular}    & \faCircleO         & \faCircle        & \faCircle         \\  
~~~\begin{tabular}[]{@{}l@{}} Backend?\end{tabular}            &    \faAdjust        & \faCircle        & \faAdjust        \\ 
~~~\begin{tabular}[]{@{}l@{}}Centralized Vetting?\end{tabular}           & \faCircle            & \faCircleO         & \faCircle         \\ 
 \midrule[1.2pt]
\multicolumn{4}{l}
{\textbf{Users}}                       \\ \midrule
~~~\begin{tabular}[]{@{}l@{}}Install-free?\end{tabular}          &   \faCircleO  & \faCircle     &  \faCircle      \\ 

~~~\begin{tabular}[]{@{}l@{}}Market?\end{tabular}          &   \faCircle   & \faCircleO           & \faCircle        \\  
~~~\begin{tabular}[]{@{}l@{}}Storage Consumption?\end{tabular}          &  High & Low          & Low        \\  
~~~\begin{tabular}[]{@{}l@{}}Update?\end{tabular}          &   Client-based  & Client-based          &  Server-based      \\  
~~~\begin{tabular}[]{@{}l@{}}Performance?\end{tabular}          &   High  & Browser-specific           & Super-app-specific        \\ 
~~~\begin{tabular}[]{@{}l@{}}Offline Loading?\end{tabular}          &   High  & Low           &  Median      \\ 
~~~\begin{tabular}[]{@{}l@{}} Register and Login?\end{tabular}          &   \faCircle  & \faCircle        &\faCircleO       \\ 

  \bottomrule
\end{tabular}
}

\caption{Comparison between different hosts and their supported apps. ``\faCircle'' represents full support; ``\faAdjust'' represents partial support; ``\faCircleO'' represents no support.}
\label{tab:cmp3apps}

  %  \vspace{-0.3in}
\end{table}

\section{Security Mechanisms in Super Apps}
\label{sec:good}
\label{sub:mechanism}

\begin{figure}[ht]
    \centering
    \includegraphics[width=1\linewidth]{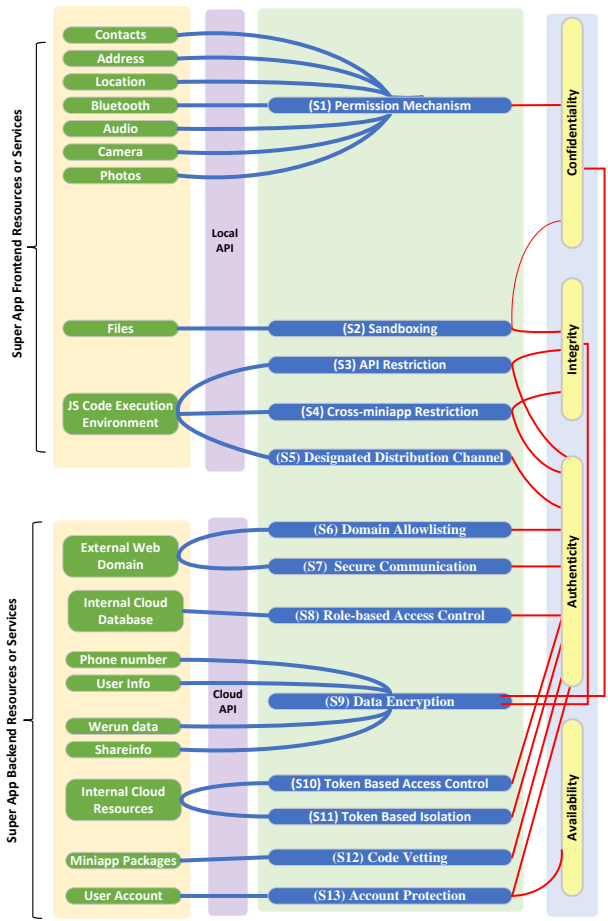}
    \caption{Relationship between the resources, security measures and the security properties}  
    \label{fig:securitymeasures}
\end{figure}

To protect the resources hosted by super apps, a series of mechanisms have been implemented to maintain the
% The security mechanisms employed by super apps are meticulously crafted to prioritize the protection of user data, maintaining the
integrity, confidentiality, and availability of sensitive information and services.
% These measures create a secure environment that benefits both users and developers alike. 
Although the exact security mechanisms may differ based on the specific implementation and platform, they  fall into two categories: front-end security measures and back-end security measures, as shown in \autoref{fig:securitymeasures}:
 
\subsection{Security Mechanisms at Front-ends}
 \label{sub:mechanismF}
Front-end security mechanisms primarily focus on safeguarding the local resources and services managed by super apps at users' devices.
 \begin{table}[]
\centering
\scriptsize
\setlength\tabcolsep{1.5pt}
\begin{tabular}{@{}lccc@{}}
\toprule
   & \begin{tabular}[c]{@{}c@{}}\textbf{Mobile OS}\\ (\textbf{Native Apps})\end{tabular} & \begin{tabular}[c]{@{}c@{}}\textbf{Web Browsers}\\ (\textbf{Web Apps})\end{tabular} & \begin{tabular}[c]{@{}c@{}}\textbf{Super Apps}\\ (\textbf{Miniapps})\end{tabular} \\ \midrule
\multicolumn{4}{l}
{\textbf{Deployed at Superapps' Frontends}}                       \\ \midrule
~~~(S1)  Permission Mechanism~\cite{scope:wechat} & \faCircle                                                                  &                                 \faCircleO                                 &                                                      \faCircle       \\

~~~(S2) Sandboxing~\cite{sandboxing:wechat} & \faCircle                                                                 &                                 \faCircle                                 &                                                         \faCircle       \\

 ~~~(S3) API Restriction~\cite{domtreeisolation:wechat,hiddenAPI2023}  & \faAdjust                                                                  &                                 \faCircleO                                  &                                                         \faCircle        \\
% ~~~(S5) Threat Monitoring  & \tickYes                                                                  &                                 \tickNo                                  &                                                         \tickYes        \\

  ~~~(S4) Cross-miniapp Allowlisting~\cite{cmrf}  & \faAdjust                                                                  &                                 \faAdjust                                  &                                                         \faAdjust       \\

  ~~~(S5)  Designated Distribution Channel~\cite{evalban:wechat} & \faAdjust                                                                 &                                 \faCircleO                                  &                                                         \faCircle        \\
 
 \midrule
 \multicolumn{4}{l}
{\textbf{Deployed at Superapps' Backends}}                       \\ \midrule

 ~~~(S6) Domain Allowlisting~\cite{networkwechat,hiddenAPI2023} & \faCircleO                                                                  &                                 \faCircleO                                  &                                                         \faCircle        \\

 ~~~(S7) Secure Communication~\cite{networkwechat}  & \faAdjust                                                                  &                                 \faCircleO                                  &                                                         \faCircle        \\
 
  ~~~(S8) Role-Based Access Control~\cite{rabc}   & \faCircle                                                                  &                                 \faCircle                                  &                                                         \faCircle        \\  
 ~~~(S9) Data Encryption~\cite{appsecretleak,wechatkey20} & \faAdjust                                                                  &                                 \faAdjust                                  &                                                         \faCircle        \\
 ~~~(S10) Token-based Services Access~\cite{wechatkey20}   & \faAdjust                                                                  &                                 \faAdjust                                  &                                                         \faCircle        \\
  
  ~~~(S11) User Token Isolation~\cite{dingdingkey,qqkey}   & \faAdjust                                                                  &                                 \faAdjust                                  &                                                         \faCircle        \\

 ~~~(S12) Code Vetting~\cite{oprules} & \faCircle                                                                  &                                 \faCircleO                                  &                                                         \faCircle        \\ 
%~~~(S13) Secure Third-Party Integration~\cite{APIlimit} & \faCircle                                                                  &                                 \faAdjust                                  &                                                         \faCircle        \\ 
  ~~~(S13) Account Protection~\cite{oprules} & \faCircleO                                                                  &                                 \faCircleO                                  &                                                         \faCircle        \\
 
  % ~~~(S14) Secure Third-Party Integration & \faCircle                                                                  &                                 \tickNo                                  &                                                         \tickYes        \\ 
 \bottomrule
\end{tabular}

\caption{Comparison of Security Mechanism. 
``\faCircle'' represents full support; ``\faAdjust'' represents partial support; ``\faCircleO'' represents no support. }  
\label{tab:comparision_sec}
 
\end{table}

\paragraph{(S1) Permission Mechanism~\cite{scope:wechat}}  Super apps typically build a strong permission mechanisms to
% to verify the identity of users and
ensure that the sensitive data (e.g., phone number, location) on users' local devices are only accessible after the users acknowledge and approve the resource access. This mechanism resembles the permission mechanisms deployed by mobile operating systems.
% only authorized access to sensitive features and data (e.g., phone number, location).  
For example,  \textsc{WeChat} has a comprehensive permission system covering a wide range of resources including user account information, social network information (e.g., group chat information), and device resources (e.g., Bluetooth). The miniapps must prompt the users of the specific type of resources required by them  prior to accessing the resources. If the users reject the resource access, the miniapps cannot access them.
% that grants or restricts access to different resources and capabilities based on user consent and privacy settings. When users interact with the miniapps, WeChat prompts them to grant specific permissions requested by the third-party miniapps. Users have the ability to accept or decline these permissions on a per-miniapp basis. 
As shown in \autoref{tab:comparision_sec}, this is very similar to the mobile permission mechanisms. \looseness=-1

\paragraph{(S2) Sandboxing~\cite{sandboxing:wechat}}  Similar to other apps that build containers to run third-party applications or OSs, super apps create controlled environments known as sandboxes as well, where each miniapp operates in independent storage spaces. This isolation ensures that the behavior of miniapps cannot directly interfere with other apps or the underlying system (e.g., writing or reading files outside the designated spaces), especially when different miniapps are given different sets of permissions (e.g., certain mini-apps may possess the privilege to access user data, while others may not), because the access to highly privileged data should not be permitted for miniapps with fewer privileges. For instance, miniapps can only access files via specific URL schemes (e.g., \texttt{wxfile} for \textsc{WeChat}), which are only stored \textit{inside} their own sandbox by super apps. As shown in \autoref{tab:comparision_sec}, mobile OSs and Web browsers have similar protections.

\paragraph{(S3) API Restriction~\cite{hiddenAPI2023}} While the super apps customize the execution engine to support 
 miniapps, it is natural for the super apps to restrict or remove APIs considered sensitive, dangerous, or vulnerable, to refrain miniapps from using them. For instance, when a miniapp invokes the payment related APIs, the super apps may check whether the developers of the miniapp have already submitted required business certificates to the platforms. Instead, developers failing to do so may not invoke the API.
% Super apps often impose API restrictions to ensure the security and integrity of their platforms. These restrictions help control access to sensitive functionalities and data, ensuring that only authorized entities can utilize certain APIs.
%One common example is restricting the \texttt{eval()} API, which executes an arbitrary string as JavaScript code, because this API may cause attackers to inject malicious script into any miniapps using this API. \CW{feels weird. I felt \texttt{eval} is not so related to Hidden API?} \ZY{I agree. eval is not related to hidden APIs} 
Meanwhile, there are ``undocumented'' APIs that are only supposed to be used by miniapps from the super app themselves, such as \texttt{searchContacts}, an API used to access the friend list of a  \textsc{WeChat} user, which is only supposed to be accessible by miniapps from \textsc{Tencent}. Super apps may verify the author of miniapps prior to the invocation to make sure the API is only accessed by expected parties.
% and accreditation process in place for developers who wish to access restricted APIs (e.g., \texttt{searchContacts} is only available for the miniapps created by the super app vendors, but not for the third-party developers).  

%

% and only miniapps from eligible developers can invoke these sensitive APIs.
% This process typically involves verifying the identity, credentials, and business legitimacy of the developers or companies. For example,
% have to submit
% Enterprise developers are required to provide their business licenses as part of the identity verification process. 

%\CW{Would this be better if we separate this section into another Security Mechanisms? like: ``Rendering Isolation'' or ``Execution Isolation''}
Aside from verifying the identity, another set of API is strictly prohibited by design to prevent exploitation. A typical example of this is that \texttt{eval()} is not accessible by miniapps, as it executes JavaScript code as strings and may be prone to injection attacks. Also, the rendering of miniapp is protected by preventing developers from accessing APIs to manipulate the Document Object Model (DOM) tree.
% The restriction of API is not limited to accessing super app services or local resources. On the other hand, the rendering of miniapps is also restricted by the super apps by DOM (Document Object Model)  tree isolation.  
% This serves as a crucial security mechanism utilized by super apps to bolster the security and isolation of miniapps within the app ecosystem. 
% Unlike traditional web apps where JavaScript can be directly executed within the DOM tree, the DOM tree for miniapps in type-B super apps in \autoref{fig:evolution} is completely separated from the JavaScript Engine. 
In these super apps, there are no APIs  manipulating DOM tree nodes to prevent miniapps from dynamically changing behavior of rendering under the similar concern of fighting against malware.
% directly executing JavaScript in DOM tree or manipulating elements in DOM tree via JavaScript is not allowed under the similar concern as restricting \texttt{eval()},
% One of the primary motivations behind DOM tree isolation is 
% to address the vulnerabilities associated with Cross-Site Scripting (XSS) attacks commonly found in traditional web app environments.By disallowing direct access to {execute JavaScript within the DOM tree}, super apps significantly reduce the risk of XSS attacks. This restriction ensures that miniapps operating within the super app cannot directly interact with or manipulate JavaScript objects present in the DOM tree environment.

\paragraph{(S4) Cross-miniapp Allowlisting~\cite{cmrf}} In order to facilitate information sharing among collaborating miniapps, super apps employ a mechanism known as cross-miniapp redirection. This mechanism enables a miniapp to redirect to another miniapp with a payload, similar to the Intent mechanism in Android. However, there is a concern that malicious miniapps may exploit this redirection and message sharing functionality to inject carefully crafted payloads to arbitrary miniapps. Consequently, super app platforms such as {\sc AliPay} have implemented an allowlist-based mechanism for miniapps to filter payload sent from unexpected miniapps. \looseness=-1

\paragraph{(S5) Designated Distribution Channel~\cite{evalban:wechat}} Super apps typically employ a strict policy to prevent hot updates.
% This policy is in place to prioritize security and stability within the app environment.  
The concept of hot update refers to the ability to update the code or functionality of an app without requiring the user to manually re-install updated packages from app stores. The hot update is prohibited in super apps because the super app needs to vet all miniapp codes before they are released to the ecosystem in case of malware. However, the hot update ability grants malicious developers capability of transforming a vetted (benign) miniapp into malware via cloud-transmitted hot update code. Hence, the ability to side-load miniapps or execute code as cloud-delivered payloads (e.g., using JavaScript interpreter libraries) are prohibited.
% While the attackers may initially submit a benign version of miniapp to the platform for it to be released to the miniapp store, the attackers can directly push update code containing malware code snippets to victims via network without having to pass the vetting mechanisms again, which poses significant threat to the super app ecosystem.
% initiated by miniapps to prevent un-supervised change of miniapp behavior that transform a ``benign'' miniapp into malware dynamically via cloud-transmitted code
 % While hot updates can offer convenience and agility in traditional app development, they introduce certain risks and challenges in the context of super apps. \looseness=-1
% \begin{mybox}[boxsep=0pt,
% 	boxrule=1pt,
% 	left=4pt,
% 	right=4pt,
%  	top=4pt,
%  	bottom=4pt,
% 	]
	
% \textbf{(Takeaway I). 
%  Super app frontends share similar security measures with mobile OSs.}
% Based on the data presented in \autoref{tab:}, it is evident that super apps share similar security measures with mobile operating systems (as most of these measures can be found in mobile OSs), but exhibit notable differences from web browsers. This distinction arises from the fact that the threat model of mobile OSs and super apps align closely. For instance, both mobile OSs and super apps are responsible for managing resources and delivering them to mobile apps or miniapps. Furthermore, both entities require code vetting, although mobile OSs generally employ less stringent vetting policies compared to super apps.
  
%  \end{mybox}

%  \vspace{-2mm}

\subsection{Security Measures at Back-ends}
 \label{sub:mechanismB}
On the other hand, back-end security measures pertain to the security controls and practices employed within the super app's server infrastructure and backend systems. These measures aim to fortify the app's core cloud functionalities, databases, submitted code, and any other components that reside on the server-side. 

\paragraph{(S6) Domain Allowlisting~\cite{networkwechat,hiddenAPI2023}}   Super apps employ a domain allowlist mechanism to effectively regulate web page access, ensuring that users are protected from downloading files or accessing content from unknown or untrusted sources.
% This security measure plays a crucial role in maintaining the integrity and safety of the super app ecosystem. 
For example, in \textsc{WeChat}, miniapps are required to use the \texttt{wx.request} function to access specific URLs. However, before granting access, \textsc{WeChat}'s vetting process verifies the requested URL against a domain allowlist. This domain allowlist is configured by developers and vetted by the platform before the miniapp is submitted, and can be viewed by users by opening the ``\texttt{About}'' page of miniapps. This verification ensures that the requested content originates from a trusted source that meets the predefined security policy. As shown \autoref{tab:comparision_sec}, we have not observed similar protections in mobile OSs and web browsers. 

\paragraph{(S7) Mandatory Secure Communication~\cite{networkwechat}}   Super apps enforce the implementation of secure communication protocols, notably HTTPS (HTTP over SSL/TLS), to establish encrypted data transmission between the miniapp and the server. For example, miniapps in \textsc{WeChat} can only use either HTTPS or WebSocket Secure (WSS) protocol, and requests sent via HTTP will be blocked by the super apps after they are released (while developers are allowed to test their communication with HTTP in development tools).

\paragraph{(S8) Role-Based Access Control (RBAC)~\cite{rabc}}  The super apps incorporate Role-Based Access Control (RBAC) mechanisms on cloud services provided by them such as cloud database to prevent the data from being accessed or overwritten by unexpected parties. While the developers have access to data stored in miniapp cloud database  provided by super apps, the privileges granted to administrators and ordinary developers are different.
% RBAC ensures that users or roles are granted appropriate access privileges based on their roles and responsibilities. 
Hence, this helps prevent unauthorized access to sensitive features and data within the miniapp.   \looseness=-1

% Access control policies are enforced at various levels, including  resources,  databases, and user data.  

%For instance, WeChat's RBAC policies, as illustrated in \autoref{tab:rbac}, outline the access rules for miniapps to retrieve their respective data. \looseness=-1

\paragraph{(S9) Data Encryption~\cite{appsecretleak,wechatkey20}}  Miniapps hosted within super apps like \textsc{WeChat} and \textsc{Baidu} often handle sensitive resources, such as user phone numbers and billing addresses. To prevent these sensitive data transmitted between miniapp front-ends and back-ends from being captured or manipulated by attackers, cryptographic protocols are employed to ensure the security of this data, relying on a master key known as the ``AppSecret''.  This key is generated by super apps, shown to developers, and supposed to be properly stored in developers' back-ends, as it is used by the super apps to verify a back-end's identity. 
% The super apps generate and manage this key, which plays a crucial role in authenticating miniapps and enabling the secure transmission of sensitive information.  
To provide more specific details, the data (such as phone numbers) is encrypted using a key called the ``session key'', and the resulting cipher is sent to the miniapp's back-end. The miniapp's back-end can then utilize the ``AppSecret'' to request the ``session key'' from the super apps, allowing it to decrypt the data.

\paragraph{(S10) Token-based Services Access~\cite{wechatkey20}}   Super apps often provide OAuth 2.0-alike protocol as a standard authentication and authorization framework for accessing their services. The protocol enables secure and delegated access to protected resources without the need to share sensitive credentials, such as usernames and passwords.
After successful authentication and authorization, the super app provides access tokens that serve as credentials for the miniapp to access specific services or resources. These access tokens are typically short-lived and can be scoped to restrict the permissions granted to the miniapp, ensuring that it only has access to the required resources.

\paragraph{(S11) User Token Isolation~~\cite{dingdingkey,qqkey}}  As mentioned in token-based service access, the cloud data access are implemented based on tokens to avoid sharing credentials via cloud communication. To prevent the user tokens from being misused or intercepted, the tokens used to access data are isolated from each miniapps. For example, when accessing Alice's user information with the data token (i.e., \texttt{DT}) under miniapp A, the back-end of miniapp A needs to submit \texttt{DT}, miniapp's ID, along with the miniapp's master key ``AppSecret''. The server checks first if the ID and master key match to ensure that the request is from an authentic source associated with miniapp A, and then send Alice's data associated with \texttt{DT}. However, even if this token is obtained by a malicious miniapp B, miniapp B cannot get the user data with \texttt{DT}, as the \texttt{DT} is only associated with miniapp A and isolated from miniapp B. As depicted in \autoref{tab:comparision_sec}, many mobile operating systems and web browsers share similar protections to those of {\bf S10} and {\bf S11}. However, these protections may not always come as standard features; they're often not a mandatory part of the system.
% \usepackage{booktabs}

\paragraph{(S12) Code Vetting~\cite{oprules}}  Super app platforms typically enforce stringent code verification and app review processes to ensure that miniapps hosted on their platforms meet certain security standards whenever miniapp packages are submitted to the platform via cloud to be released to the miniapp market. This helps mitigate the risks associated with malicious or vulnerable apps being distributed to users (e.g., miniapps with leaked AppSecret). This approach is akin to the code verification and review practices implemented by mobile App Markets (e.g., Apple AppStore and Google Play). \looseness=-1

% \paragraph{(S13) Secure Third-Party Integration~\cite{APIlimit}}   Secure third-party integrations in super apps involve careful measures to ensure the integrity and security of the miniapps and their associated plug-ins.  These plug-ins are essentially packed with the same structures as miniapps do, but they provide a set of encapsulated services and UI layouts and are designed to be integrated inside normal miniapps (e.g., payment plugins can be integrated inside shopping miniapps, where payment plugins alone are not visible from users nor can be executed independently \AY{may need to confirm this}). 
% % Secure third-party integrations in super apps involve careful measures to ensure the integrity and security of miniapps and their associated plug-ins. 
% These miniapp plug-ins operate independently from the host miniapp, with restricted access to privileged APIs and no direct access to sensitive data. Also, the plugins are executed in a passive manner, meaning that they typically can only receive data or be invoked, instead of send data to miniapps or actively invoke miniapp services. Meanwhile, these plugins are hosted in a special plugin market which is only visible to developers. Whenever a developer integrates a plugin, he/she must first apply and seek approval from the developer of plugin. This mechanism is enforced to prevent abuse from untrusted parties.
% The super app ecosystem implements integrity checks for plug-ins, validating their authenticity and preventing the integration of malicious or tampered code. 

\paragraph{(S13) Account Protection~\cite{oprules}} Super apps implement various mechanisms to protect user accounts from being stolen, as well as vetting mechanisms on the accounts to prevent malicious developers from obtaining an account at a low cost to exploit the platform.
% these accounts often involve sensitive privacy and financial related services and data. These mechanisms include account verification and supervision. 
Upon registration, users are required to provide personal information such as resident ID and phone numbers to the platform. When registering as developers, more information may be required, such as commercial license for enterprise developers. Then, whenever a user uses super apps, or a developer accesses the developer tool, verification is enforced via log in. Meanwhile,
% Super apps often require users to verify their identities during the registration process. This may involve providing personal information, phone number verification, or linking to social media accounts. Account verification helps ensure that users are genuine and enhances the overall security of the platform. Meanwhile, 
super apps have mechanisms in place to suspend or ban user accounts if suspicious or malicious activity is detected, such as modifying network traffic using software or engaging in fraudulent behavior.

\subsection{Summary of Security Assumption}
\label{sub:assumption}
% The super apps prioritize the protection of user data by enforcing security mechanisms at front-end (the super app on end users' devices) and back-end (the super app cloud server) against malicious developers and users from exploiting the ecosystem. 
The security mechanisms enforced in super apps are designed based on different security assumptions which can be categorized into three categories, including the ``Adopted'' assumption, ``Isolated'' assumption, and ``Vetted'' assumption.  \looseness=-1

  \paragraph{(A1) The ``Adopted'' Assumption} As the super apps share similarity with mobile and web computing paradigms, four out of 13 mechanisms (\textbf{S1}, \textbf{S2}, \textbf{S7}, and \textbf{S8}) introduced in this section are adopted from traditional paradigms with little customization. As these mechanisms are adopted from sophisticated paradigms directly from traditional platforms, the security assumption is assumed to be as strong as traditional platforms. For instance, while permission mechanisms are built upon Android system to manage super app specific data, the mechanisms to require users to acknowledge and approve data access is essentially the same as permission mechanisms in Android systems. Similarly, the idea of RBAC and Sandboxing are adopted from OS resource management model, and Secure Communication (HTTPS requirement) is directly adopted from web computing.
  
 \paragraph{(A2) The ``Isolated'' Assumption} Super apps incorporate various mechanisms to ensure that resource access from third parties, particularly untrusted or non-authentic entities, is appropriately isolated. Among the 13 mechanisms implemented, four are associated with this assumption (\textbf{S3}, \textbf{S9}, \textbf{S10}, and \textbf{S11}). These mechanisms include isolating privileged APIs to prevent non-super-app miniapp developers from accessing them (\textbf{S3} API restriction). Additionally, the data access is isolated from non-original miniapp developers through the enforcement of master keys and tokens at a per-miniapp level (\textbf{S9}, \textbf{S10}, and \textbf{S11}). By enforcing these mechanisms and ensuring that each miniapp can only access its own resources, the security of data access is intended to be effectively safeguarded. \looseness=-1
 
 \paragraph{(A3) The ``Vetted'' Assumption} Since super apps now feature  exclusive miniapp market where all executable miniapps have to be submitted and vetted by the super apps, it is assumed that the super app platform will vet and identify potential malicious behaviors of third-party miniapps. These vetting covers the rest five out of 13 mechanisms (\textbf{S4}, \textbf{S5}, \textbf{S6}, \textbf{S12}, and \textbf{S13}), including vetting of miniapp communication between front-ends (\textbf{S4}) and back-ends (\textbf{S6}), control on miniapp releasing and distribution (\textbf{S5} and \textbf{S12}), as well as  vetting and supervision on developer accounts (\textbf{S13}).

\ignore{
\begin{enumerate}[label={(\bfseries G\arabic*)}]

\item \textbf{Monetizing Popularity:} The miniapp paradigm allows super apps to offer versatile and convenient third-party services chosen by users, attracting more users and utilizing the large user base for profitable activities like advertising and paid media streaming. In 2017, social app giants Tencent (China) and Kakao (Korea) introduced miniapps, following a decline in active users in the social networking market after a surge in social apps~\cite{socialfatigue}. Despite this, Tencent's social advertising revenue increased by 61\% to \$1.4 billion in 2018 after introducing miniapps in WeChat~\cite{miniapprevenue}. The introduction of new advertising methods (miniapps) and additional resources contributed to this growth, as mentioned in the annual report. By 2022, social advertising revenue doubled to \$3.1 billion~\cite{miniapprevenue2}. 

\item \textbf{Reducing Financial Costs:} 
The miniapp paradigm offered by super app platforms allows small business owners to access cloud resources conveniently and at low cost. They can utilize cloud databases and functions provided by the platform to process orders and account information for a large user base. This eliminates the need for expensive server rental and database maintenance. Additionally, super app platforms offer affordable paid-for cloud services, such as Face Recognition, freeing developers from implementing them independently. As a result, small businesses can integrate miniapps via the cloud with minimal costs and provide convenient services like remote ordering to users.

\item \textbf{Mitigating Security Risks:} 
Enabling third-party developers to provide services on super app platforms introduces potential security risks, with attackers exploiting user data and the platform itself. To combat this, super app platforms require all third-party miniapps to undergo platform vetting before reaching end users. Developers must register for accounts, providing personal or enterprise information for evaluation. After developing a miniapp, they submit it for review.  The super apps restrict users from loading miniapps from unofficial sources, allowing access only through the built-in miniapp store or shared links with icons and descriptions. These measures give the platforms control over third-party codes and mitigate malicious behaviors, ensuring user security within the miniapp ecosystem.

\item \textbf{Reducing Technical Barriers:}
Various techniques, such as templates and off-the-shelf libraries, within this ecosystem serve to lower the barrier for developers. For example, even business owners without extensive software development experience can have their own miniapps published in super apps using miniapp templates. With a template, a shop owner can purchase one from a third-party, upload product information, and have the miniapp released on their behalf. The popularity and feasibility of miniapp templates stem from the similarities in functionality among miniapps, which are closely tied to the features of the super app. For example, a significant portion of shopping miniapps (38\% in WeChat and 78\% in Baidu) exist due to the built-in payment capabilities of these super apps. The prevalence of similar needs among miniapps has created a thriving secondary market for miniapp templates~\cite{cmrf}.

\item \textbf{Improving User Experience:} In a seamless experience within a single super app and account, miniapps are widely adopted in restaurants and shops, particularly those integrated into payment-supporting super apps like AliPay and WeChat. Imagine Alice walking into a flower shop, selecting her favorite flowers, and realizing she left her wallet behind. Instead of a long trip back home, she simply scans the QR code on the shop's wall and pays for her flowers. While the flowers are being prepared, she proceeds to place an order at a nearby restaurant using her phone number and the in-app purchasing feature. To pass the time, she launches a miniapp game to enjoy with her friends. Once her flowers are ready, she receives an SMS notification confirming the order's completion. She casually walks across the street to collect her dinner and heads home, all without switching between different mobile apps, re-registering accounts, or setting up payment passwords.

\end{enumerate}

There are also other advantages when the super apps become ``operating systems'', and we cannot enumerate them all. For example, this also encourages innovation and competition among developers, resulting in a rich selection of apps and services for users to choose from; With mobile super apps, users typically have a single account that grants them access to various services within the platform. This eliminates the hassle of creating and managing multiple accounts for different apps.  Super apps often integrate secure and convenient payment solutions, allowing users to make transactions seamlessly within the app. This eliminates the need for multiple payment apps or cash transactions, providing users with a hassle-free payment experience. 

}

\section{Security Threats and Root Causes}
\label{sec:bad}

\subsection{Threats at Front-ends}
% Threats against front-ends of mini-apps and super apps can pose significant risks to the security and integrity of these applications. Poorly designed or implemented front-ends may lack proper security controls, making them vulnerable to attacks. Inadequate use of secure coding practices, weak authentication mechanisms, or flawed session management can create openings for attackers to compromise the front-end. To be more specific: 

\paragraph{(T1) Data Leakage Due to Flawed Permission~\cite{lu2020demystifying}} While super apps build permission mechanisms upon underlying systems, the permission enforced by super apps are not necessarily a strict superset of the underlying system. More specifically, there are inconsistencies on permission required between miniapp APIs and system APIs. For instance, Lu \textit{et al.}~\cite{lu2020demystifying} discovered that while Android system requires applications to be granted location permission (\texttt{ACCESS\_COARSE\_LOCATION} and \texttt{ACCESS\_FINE\_LOCATION}), accessing Wi-Fi List via \texttt{wx.getWifiList} in \textsc{WeChat} did not require similar super app permission. % by the time of 2020. 
While this vulnerability has been fixed by mainstream super app platforms by enforcing miniapps to be granted \texttt{location} permission, the mismatch between permission management and super apps still may pose a threat, as the underlying system's permission management is constantly being updated, creating potential mismatches. 

% In a mini-app context, attackers may create fake interfaces that closely resemble the legitimate mini-app's user interface, tricking users into entering their personal information. These fraudulent interfaces may prompt users to log in, provide account details, or submit sensitive data under the pretense of some urgent or enticing reason. The purpose of a phishing attack in a mini-app is to collect users' confidential information, which can be subsequently misused for various malicious activities, such as identity theft, unauthorized access to accounts, or financial fraud.

\paragraph{(T2) Data Leakage Due to Cross-platform Vulnerabilities~\cite{crossplatform2023}} Supporting multiple platforms (such as Android and iOS) introduces challenges in maintaining consistent security measures, as divergent security policies, discrepancies of implementation, and inconsistent security updates across platforms can create gaps that attackers may exploit. 
% Moreover, certain miniapp APIs are platform-specific, with some only available on specific platforms. 
For instance, sensor APIs like accelerometer, compass, and gyroscope are typically limited to mobile platforms such as Android and iOS. However, the difference in APIs may go beyond compatibility, introducing security vulnerabilities.
% While most API differences may result in compatibility issues, some can introduce security vulnerabilities. 
One such example is the \texttt{makeBluetoothPair} API, which facilitates Bluetooth device authentication. The absence of this API on platforms like iOS prevents Bluetooth devices from distinguishing between trusted and untrusted devices, thereby opening the door for potential Man-in-the-Middle attacks.\looseness=-1

\paragraph{(T3) Data Leakage via Hidden APIs~\cite{hiddenAPI2023}} While abundant documentation have been provided to developers about functionalities and APIs, not all APIs are documented. These ``hidden'' APIs are designed for internal use (e.g., miniapp developers from Tencent only in \textsc{WeChat}), but may potentially enable miniapps to bypass restrictions and gain unauthorized access.  For example, in WeChat, a malicious miniapp can exploit the hidden API to access arbitrary malicious contents without being detected by the super apps or secretly download and install harmful Android apps. These malicious apps have the potential to steal sensitive user information. The usage of hidden APIs such as ``\texttt{captureScreen}'' to capture screenshots, ``\texttt{getLocalPhoneNumber}'' to steal the user's phone number, and ``\texttt{searchContacts}'' to extract the user's contact information. 
%These examples underscore the risks associated with hidden APIs and their potential for privacy breaches and data theft.

% \paragraph{(T2) Data Leakage via Collusion~\cite{taintmini2023}} In an ideal scenario, each miniapp should only have explicit user-granted permissions and access to the corresponding sensitive resources. However, similar to how Android apps can communicate through intents, miniapps also have the capability to cross-communicate. This cross-miniapp communication enables the completion of complex tasks but also introduces the risk of collusion attacks, similar to those found in traditional mobile apps. For instance, when using a bike-renting miniapp, users first select a bike and then pass the rent fee to a payment miniapp to complete the transaction. Unfortunately, the unrestricted nature of this communication means that a miniapp with access to sensitive data can potentially leak that data to other miniapps without the user's awareness. \looseness=-1

\paragraph{(T4) Data Injection via Cross-miniapp Channel~\cite{cmrf}} 
Miniapps, due to their limited size, often rely on communication with other miniapps to accomplish complex tasks. For instance, a shopping miniapp may need to exchange information, such as the order ID and price, with a payment miniapp to facilitate a purchase. As such, this cross-communication becomes a vital means for miniapps to enhance their functionalities and collaborations. However, if a receiver miniapp fails to perform proper checks on the sender's appId, it can result in a new form of attack called Cross-Miniapp Request Forgery (CMRF), where a malicious miniapp can send arbitrary payloads via cross-miniapp channel. These payloads may be fabricated order data to trigger shopping-for-free, or login data to intrude into an arbitrary user's account.

% \usepackage{multirow}

\begin{table*}[]
\scriptsize
\centering
\setlength{\belowrulesep}{0pt}
\setlength{\aboverulesep}{0pt}
\setlength{\tabcolsep}{2.5pt}
\begin{adjustbox}{angle=0}
\begin{tabular}{cc|c|cc|ccc|cc|cccc|c|ccc}
\toprule
\multicolumn{8}{c|}{\textbf{Security Mechanism Analysis}}& \multicolumn{6}{c|}{\textbf{Security Threat Analysis}}& \multicolumn{4}{c}{\textbf{Security Impact Analysis}} \\\hline
\multicolumn{2}{c|}{\multirow{2}{*}{\textbf{Security Mechanism}}} & \multicolumn{3}{c|}{\textbf{At}} & \multicolumn{3}{c|}{\textbf{Assumption}} &\multicolumn{2}{c|}{\multirow{2}{*}{\textbf{Threat ID}}} & \multicolumn{4}{c|}{\textbf{Root Cause}}   & \textbf{Privileged}&\multicolumn{1}{c|}{\textbf{Vetting}}  & \multicolumn{2}{c}{\textbf{Data Issue}}   \\ \cline{3-8}\cline{11-14}\cline{17-18}
\multicolumn{2}{c|}{}&\textbf{\#}&\textbf{F}&\textbf{B}&\textbf{A}&I&\textbf{V}&\multicolumn{2}{c|}{}&\textbf{Comp.}&\textbf{Impl.}&\textbf{Trust}&\textbf{Vetting}&\textbf{Access}&\multicolumn{1}{c|}{\textbf{Bypass}}&\textbf{Injection}&\textbf{Leakage}\\\hline
 % && \cellcolor[HTML]{FFFFFF}\textbf{\#} & \cellcolor[HTML]{FFFFFF}\textbf{F} & \cellcolor[HTML]{FFFFFF}\textbf{B}   & \cellcolor[HTML]{FFFFFF}\textbf{A} & \cellcolor[HTML]{FFFFFF}\textbf{I} & \cellcolor[HTML]{FFFFFF}\textbf{V} &&&\multicolumn{1}{c|}{\cellcolor[HTML]{FFFFFF}\textbf{Comp.}} & \cellcolor[HTML]{FFFFFF}\textbf{Impl.} & \cellcolor[HTML]{FFFFFF}\textbf{Trust} & \multicolumn{1}{c|}{\cellcolor[HTML]{FFFFFF}\textbf{Vetting}} & \cellcolor[HTML]{FFFFFF}\textbf{Access}  & \multicolumn{1}{c|}{\cellcolor[HTML]{FFFFFF}\textbf{Bypass}} & \multicolumn{1}{c|}{\cellcolor[HTML]{FFFFFF}\textbf{Inj.}} & \cellcolor[HTML]{FFFFFF}\textbf{Leak.} \\
\rowcolor[HTML]{CFE2F3} 
S1& \begin{tabular}[c]{@{}c@{}}Permission \\ mechanism\end{tabular}&\ding{172}   & \tickYes  &  & \tickYes   &   &  & T1   & \begin{tabular}[c]{@{}c@{}}Flawed \\Permission\end{tabular}   &\begin{tabular}[c]{@{}c@{}}Permission\\ Management\end{tabular} & & & & Data & &  & \tickYes\\\hline
\rowcolor[HTML]{CFE2F3} 
S2& Sandboxing & \ding{172}   & \tickYes  &  & \tickYes&   &  & T2   & \begin{tabular}[c]{@{}c@{}}Cross-platform\\ Vulnerability\end{tabular}  & \begin{tabular}[c]{@{}c@{}}Resource\\ Management\end{tabular}  & &  & & Data & && \tickYes \\\hline
\rowcolor[HTML]{FCE5CD} 
S3& API Restriction& \ding{173}   & \tickYes  &  &  & \tickYes &  & T3   & \begin{tabular}[c]{@{}c@{}}Hidden API \\ Access\end{tabular}&& Missing & & & Service  & && \tickYes \\\hline
\rowcolor[HTML]{D9D2E9} 
S4& \begin{tabular}[c]{@{}c@{}}Cross-miniapp \\ Allowlisting\end{tabular}  & \ding{174}   & \tickYes  &  &  &   & \tickYes& T4   & \begin{tabular}[c]{@{}c@{}}Cross-miniapp \\ Injection\end{tabular} && & Miniapp& & Miniapp & & \tickYes  &  \\\hline
\rowcolor[HTML]{D9D2E9} 
S5& \begin{tabular}[c]{@{}c@{}}Designated \\ Distribution Channel\end{tabular} & \ding{175}   & \tickYes  &  &  &   & \tickYes& T5   & \begin{tabular}[c]{@{}c@{}}Post-vetting\\ Hot Update\end{tabular}  && &  & \multicolumn{1}{c|}{Post-vetting}& Service & \tickYes   &&  \\ \toprule
\rowcolor[HTML]{D9D2E9} 
S6& Domain Allowlisting& \ding{176}   && \tickYes&  &   & \tickYes& T6 &\begin{tabular}[c]{@{}c@{}}Identity Confusion\end{tabular}&&&Miniapp&&Service&\tickYes&&\\\hline
\rowcolor[HTML]{CFE2F3} 
S7& \begin{tabular}[c]{@{}c@{}}Secure \\ Communication\end{tabular}& \ding{176}   && \tickYes& \tickYes&   && \multicolumn{2}{c|}{-}  & \multicolumn{4}{c|}{-}  &- &\multicolumn{3}{c}{-}  \\\hline
\rowcolor[HTML]{CFE2F3} 
S8& \begin{tabular}[c]{@{}c@{}}Role Based \\ Access Control\end{tabular}   & \ding{177}   && \tickYes& \tickYes&   &  & \multicolumn{2}{c|}{-}  & \multicolumn{4}{c|}{-}    &-& \multicolumn{3}{c}{-}  \\\hline
\rowcolor[HTML]{FCE5CD} 
S9& Data Encryption& \ding{177}   && \tickYes&  & \tickYes &  & T7   & \begin{tabular}[c]{@{}c@{}}Master Key\\ Misuse\end{tabular} && & Developer& & Data & && \tickYes\\\hline
\rowcolor[HTML]{FCE5CD} 
S10   & \begin{tabular}[c]{@{}c@{}}Token-based \\ Access Control\end{tabular}  & \ding{178}   && \tickYes&  & \tickYes &  & T8   & \begin{tabular}[c]{@{}c@{}}Abused \\API Token\end{tabular}   && & Developer& & Data & && \tickYes\\\hline
\rowcolor[HTML]{FCE5CD} 
S11   & User Token Isolation   & \ding{178}   && \tickYes&  &\tickYes &  & T9 & \begin{tabular}[c]{@{}c@{}}Weak Token \\ Isolation\end{tabular} && Weak& & & Data & && \tickYes\\\hline
\rowcolor[HTML]{D9D2E9} 
S12   & Code Vetting   & \ding{179}   && \tickYes&  &   & \tickYes&T10&\begin{tabular}[c]{@{}c@{}}Evasive Malware\end{tabular}&&&&Intra-vetting&Data&\tickYes&&  \\\hline
\rowcolor[HTML]{D9D2E9} 
S13   & Account Protection & \ding{180}   && \tickYes&  &   & \tickYes&  \multicolumn{2}{c|}{-}  & \multicolumn{4}{c|}{-}&- & \multicolumn{3}{c}{-} 
\\\bottomrule
\end{tabular}

\end{adjustbox}

\setlength{\tabcolsep}{6.8pt}
\begin{tabular}{c|c|c|c|c|c|c|c|c|c}

\multicolumn{3}{c}{}\\
\toprule
\multicolumn{10}{c}{\textbf{Explaination of the Abbreviations}}\\\hline
\multicolumn{3}{c|}{\textbf{Mechanism Enforcement~(\S\ref{sub:mechanismF},~\S\ref{sub:mechanismB})}}&\multicolumn{3}{c|}{\textbf{Security Assumption~(\S\ref{sub:assumption})}}&\multicolumn{4}{c}{\textbf{Root Cause Analysis~(\S\ref{sub:rootcause})}}\\ \hline
     \textbf{\#}&\textbf{F}&\textbf{B}&{\cellcolor[HTML]{CFE2F3} \textbf{A}}& {\cellcolor[HTML]{FCE5CD} \textbf{I}}&{\cellcolor[HTML]{D9D2E9} \textbf{V}}&\textbf{Comp.}&\textbf{Impl.}&\textbf{Trust}&\textbf{~~~~Vetting~~~~} \\ \hline

     \begin{tabular}[c]{@{}c@{}}ID of edge \\in \autoref{fig:paradigm}\end{tabular}&\begin{tabular}[c]{@{}c@{}}Enforced\\At Front-end\end{tabular}&\begin{tabular}[c]{@{}c@{}}Enforced\\At Back-end\end{tabular}&\begin{tabular}[c]{@{}c@{}}\cellcolor[HTML]{CFE2F3} ``Adopted''\\\cellcolor[HTML]{CFE2F3}Assumption\end{tabular}& \begin{tabular}[c]{@{}c@{}}\cellcolor[HTML]{FCE5CD} ``Isolated''\\\cellcolor[HTML]{FCE5CD}Assumption\end{tabular}&\begin{tabular}[c]{@{}c@{}}\cellcolor[HTML]{D9D2E9} ``Vetted''\\\cellcolor[HTML]{D9D2E9}Assumption\end{tabular} &\begin{tabular}[c]{@{}c@{}}Compatibility \\ Issue\end{tabular}&\begin{tabular}[c]{@{}c@{}}Implementation \\ Issue\end{tabular}&\begin{tabular}[c]{@{}c@{}}Trust Model\\Issue\end{tabular}&\begin{tabular}[c]{@{}c@{}}Vetting\\Issue\end{tabular}\\
     \bottomrule
\end{tabular}
\caption{Summary of protection mechanisms, security assumptions, and threats due to violation of assumptions. The rows are colored based on categories of security assumption.}
\label{tab:allproblems}
%\vspace{-0.2in}
\end{table*}

\paragraph{(T5) Vetting Bypass via Post-vetting Hot Update~\cite{evalban:wechat}} The use of hot updates by malicious miniapps to bypass vetting in super apps poses a significant security concern. Hot updates refer to the ability to dynamically update or modify the code of an application without requiring a full version update or undergoing the typical vetting and approval processes. While certain known APIs like \texttt{eval()} and the use of \texttt{VM} library~\cite{vm} is prohibited in super app platforms to prevent dynamically loaded or interpreted code, the malicious actors may still exploit hot update restriction by implementing and integrating an interpreter on their own~\cite{jsjs}, as JavaScript is essentially an interpreted language, and the code does not have to be compiled.

\subsection{Threats at Back-ends}

% Threats against the back-ends of mini-apps and super apps can stem from various high-level root causes. Back-ends with inadequate authentication and authorization mechanisms are prone to unauthorized access. Weak key management policies, improper token management, or flawed access control mechanisms can be exploited by attackers to gain unauthorized privileges or access sensitive data.

\paragraph{(T6) Vetting Bypass via Identity Confusion~\cite{zhang2022identity}} Security concerns arise in the super app ecosystem regarding access control for privileged APIs. Existing super apps lack atomic identities, violating the principle of least privilege. For example, Zhang \textit{et al.}~\cite{zhang2022identity} discovered that an unprivileged miniapp may contain a privileged web domain, a privileged miniapp ID may include unprivileged third-party web domains, or an unprivileged miniapp may gain privileged capabilities. This situation creates a vulnerability known as ``identity confusion''. Adversaries can exploit this vulnerability by manipulating their own identity to masquerade as entities with granted permissions, thereby confusing the super app during the identity verification process. This identity confusion vulnerability, if present in an app-in-app ecosystem, poses a significant security risk that needs to be addressed.\looseness=-1
 
\paragraph{(T7) Data Leakage Due to Key-Misuse~\cite{appsecretleak}} As discussed in \textbf{S9}, 
robust cryptographic protocols rely on a master key called the ``AppSecret'' generated and managed by the super app to ensure the security of data, which is essential for authenticating miniapps and securely transmitting sensitive information. However, if this key is leaked in the front-end of the miniapps, it can have severe consequences for both developers and users. Upon extracting the miniapp packages and harvesting these AppSecrets, attackers can now obtain the encryption key and manipulate the encrypted data for harmful activities, including account hijacking (unauthorized access to others' accounts), promotion abuse (exploiting promotional activities for personal gain), and service theft (utilizing others' paid services without authorization). 
    
\paragraph{(T8) Data Leakage Due to Abused Token~\cite{appsecretleak}} Super apps and web apps alike use OAuth 2.0 to protect user data accessed by miniapps. On top of that, API tokens are also generated from AppSecrets for developers to invoke various cloud services such as Optical Character Recognition (OCR). However, as the AppSecrets and tokens may be leaked by developers to miniapp front-ends, which can be intercepted and abused by attackers, these cloud data and services may be provided to unexpected malicious parties. Moreover, many cloud services provided by super app platforms are not free, henceforth the abuse of API tokens may further inflict financial losses to victim developers who leaked the keys.
% Instead of directly providing phone numbers, apps like JingDong and WeChat generate data tokens. These tokens, along with app ID and app secret, are sent to the miniapp's backend, which then requests the user's phone number from the super app's cloud server. 

\paragraph{(T9) Data Leakage Due to Weak Token Isolation} The tokens used to access user data are designed as an access control mechanism as the token is isolated at per-user-per-miniapp basis, which indicates that a token is only associated with a specific user under a specific miniapp, and this is why the AppSecret of miniapps has to be submitted to super app cloud to get the user data. However, the implementation of isolation is not the same across platforms.
% This per-user-per-miniapp isolation ensures that even if a malicious user obtains the token, they cannot manipulate or access the phone number. However, not all super apps enforce this isolation. 
For example, in super app {\sc DingTalk}, upon fetching a data token from a user, the token can be used by different miniapps under the same developer. However, this poses a threat of colluding attack, as a low-privilege miniapp may receive data token from a high-privilege miniapp, but still can fetch the user data protected by the data token from the platform.

\paragraph{(T10) Vetting Bypass by Evasive Malware} With the vetting mechanism enforced, malicious miniapps seek to bypass the vetting, which results in an endless arms race between vetting mechanisms and evasive malware. For example, malicious developers may obfuscate their code to make it harder for the vetting process to detect their true intentions, or deploy dynamic loading techniques. Moreover, as the super apps involve social network among users, it has become a favorable target for launching phishing and scam malware to be rapidly distributed among users' social networks. Beyond that, there are more types of vaguely-defined malware, such as miniapps cheating the popularity by forcing users to forward miniapps to friends, because miniapps with advertisement embedded receive more profit based on higher count of clicks. These cheating behaviors include forcing users to share, faking as authentic miniapps, or acting as miniapp portal, because user will then use the portal miniapp to enter all other miniapps, where portal miniapps automatically get clicked and thus receive more profit from advertisers.  \looseness=-1

\subsection{Root Cause Analysis}
\label{sub:rootcause}
In this paper, we identified 10 threats that violated the security assumptions on the security mechanisms enforced by super app platforms, which can be categorized into compatibility, implementation, trust mode, and vetting issue as follows.   \looseness=-1

\paragraph{Compatibility Issue} As super apps execute upon different operating systems (Android, Windows, iOS, etc.), while the super apps may have implemented security mechanisms like sandboxing, the security property still are affected by how the underlying systems manage the accessible resources. Thus, issues in permission mechanisms (\textbf{T1}) and resource management (\textbf{T2}) may still create vulnerabilities in super apps. Hence, the super apps need to pay attention to discrepancies across underlying systems, and to implement additional protection if needed.

\paragraph{Implementation Issue} While super apps implemented various isolation mechanisms to perform access control on miniapps accessing resources, the implementation may suffer from missing (\textbf{T3}, missing or programmatically API isolation) or weak (\textbf{T11}, weak isolation of tokens) spots that eventually weakens or even voids the mechanism. Hence, super app platforms need to ensure that the implementation is following standardized practices and cover comprehensive aspects of the ecosystem.

\paragraph{Trust Model Issue} While security mechanisms have been implemented to enable access control on untrusted parties, some security assumptions on what can be trusted still contain flaws. For example, while developer identities are vetted by the super app platforms, the platforms still should not trust the developers too much that they will never leak sensitive tokens, as the developers may not be aware of the importance of leak these tokens by mistake (\textbf{S9}, \textbf{S10}). Also, the super apps should not place excessive trust on miniapps even if they are vetted, as they may contain vulnerabilities not covered by vetting (\textbf{T4}, \textbf{T6}).

\paragraph{Vetting Issue} The other aspect of the trust issue is that even though super apps adopt strict vetting mechanism and controls the distribution of miniapps, the vetted miniapps still should not be overly trusted. During vetting, malware may actively work against the vetting mechanism, where vetting may not discover all malicious behaviors (\textbf{T12}). After vetting, malware still may be able to dynamically load malicious code without being identified by the platforms (\textbf{T5}).

\begin{table}[]
\centering
\scriptsize
\setlength\tabcolsep{2pt}
\begin{tabular}{@{}lccc@{}}
\toprule
   & \begin{tabular}[c]{@{}c@{}}\textbf{Mobile OS}\\ (\textbf{Native Apps})\end{tabular} & \begin{tabular}[c]{@{}c@{}}\textbf{Browsers}\\ (\textbf{Web Apps})\end{tabular} & \begin{tabular}[c]{@{}c@{}}\textbf{Super Apps}\\ (\textbf{Miniapps})\end{tabular} \\ \midrule
\multicolumn{4}{l}
{\textbf{Threats against Frontends}}                       \\ \midrule
~~~(T1)  Flawed Permission  &  \faCircle                                                                   &     \faCircleO                                                          &                                                        \faCircle         \\

~~~(T2)  Cross-platform Vulnerability &                                                             \faCircleO      &          \faCircleO                                                      &                                                        \faCircle      \\
~~~(T3)  Hidden API Access  & \faCircle                                                                  &                               \faCircle                               &                                                        \faCircle      \\
 ~~~(T4)  Cross-miniapp Injection & \faCircle                                                                  &                                 \faCircle                                 &                                                         \faCircle      \\
 ~~~(T5) Post-vetting Hot Update & \faAdjust                                                                  &                                 \faCircleO                                &                                                         \faCircle       \\
 
  % ~~~(T6) Data Leakage (Key Misuse)  & \tickNo                                                                  &                                 \tickNo                                  &                                                         \tickYes        \\
  %  ~~~(T7) Data Leakage (Abused Token) & \tickYes                                                                  &                                 \tickYes                                 &                                                         \tickYes        \\
 \midrule
 \multicolumn{4}{l}
{\textbf{Threats against Backends}}                       \\ \midrule
 
 ~~~(T6) Identity Confusion & \faCircleO                                                                  &                               \faCircleO                               &                                                         \faCircle       \\
 ~~~(T7) MasterKey Misuse & \faAdjust                                                                  &                                 \faAdjust                                 &                                                         \faCircle        \\
 ~~~(T8)  Abused Token  & \faAdjust                                                                   &                                 \faAdjust                                &                                                         \faCircle       \\
 
~~~(T9) Weak Token Isolation &    \faAdjust                                                                   &                                 \faAdjust                                &                                                         \faCircle        \\
 ~~~(T10)  Evasive Malware & \faCircle                                                               &                                 \faCircleO                                  &                                                         \faCircle        \\
 %  ~~~(T12) Vetting Bypass via Hot Update   & \tickYes                                                                  &                                 \tickNo                                 &                                                         \tickYes        \\  

 % ~~~(T13) Vetting Bypass via Content Switching  & \tickYes                                                                  &                                 \tickNo                                  &                                                         \tickYes        \\ 
 %  ~~~(T14) Single Point of Failure & \tickYes                                                                  &                                 \tickYes                                &                                                         \tickYes        \\ 
  % ~~~(S14) Secure Third-Party Integration & \tickNo                                                                  &                                 \tickNo                                  &                                                         \tickYes        \\ 
 \bottomrule
\end{tabular}
\caption{Comparison of Threats.
\faCircle: the threat works against the paradigm; \faAdjust: the threat works against some of the implementations; \faCircleO: the threat has not been discovered in the paradigm.}
\label{tab:threats}
\end{table}

\section{Lessons Learned}
\label{sec:lessons}
In this paper, we systematically investigated super app platforms, summarized 13 protection mechanisms (5 at front-end, 8 at back-end), and identified 10 security threats due to violation of three families of security assumptions. By examining the observations summarized in \autoref{tab:allproblems}, and comparing the threats against traditional paradigms in \autoref{tab:threats}, in this section, we present 5 lessons learned from the threats in super apps as follows.

\subsection{The Inherited Debt of Super Apps}

\paragraph{(L1) The old is new again} As shown in \autoref{tab:threats}, a majority of threats in super apps can be found in traditional paradigms, such as hidden APIs threats which broadly existed in mobile and web apps. However, these ``old'' problems have become new again under the context of super app platforms. For instance, while cross-app injection existed in Android systems, the systems do not necessarily have to be responsible for the attack, as the attack happens between two apps that are not hosted by the systems themselves (they could be downloaded from third-party websites), and thus it is partly the users' responsibility to beware of unknown apps. Hence, the Android system informs users of apps from unknown sources but does not restrict their installation. Super apps, however, exclusively host the miniapp packages, and thus have more liability on miniapps targeting this vulnerability, especially when these miniapps access privacy-sensitive data hosted by super apps.

% the cross-miniapp injection

\paragraph{(L2) Every coin has two sides} While super apps have combined the benefits of executing versatile web apps with powerful native functionalities, super apps also face threats generated from both paradigms. These include problems derived from mobile paradigms (e.g., permission, cross-app channel, hot update, etc.) and from web paradigms (token-related vulnerabilities, API restrictions, and filterings). Hence, the super app platforms have to prudently design the protection mechanism to safeguard the user data stored in the ecosystem.

\subsection{Violation of Security Assumption}
As shown in \autoref{tab:allproblems}, we have summarized all the security mechanisms, assumptions, and threats due to the violation of these assumptions. When looking into mechanisms based on ``Adopted'', ``Isolated'', and ``Vetted'' assumptions, we may draw different observations on the root cause of the threats.

\paragraph{(L3) Compatibility is an issue} Mechanisms under the Adopted assumption suffer from compatibility issues and trust model change. This includes permission mechanisms which suffer from super app's inconsistency on permission set compared with the underlying system, and sandboxing issues suffered from different mechanisms across platforms (especially between super apps on desktop systems and mobile systems). Hence, the super apps have to check whether the protection mechanisms are implemented to offer the same level of protection on all potential systems the super apps will be executed on.

\paragraph{(L4) Single leak sinks the ship} Mechanisms under the isolated assumption suffer from weak or missing implementation, which voids the isolation mechanism. We have identified four isolation mechanisms, with which one is enforced on accessible API and three is enforced on cloud communication. From the table, we observe that two threats (\textbf{T3}, \textbf{T9}) are due to super app's implementation. 
% One is because super apps do not implement programmatic isolation on APIs (missing), and the other one is because certain super apps implement token isolation in a less strict way. 
In terms of the threats due to careless developers, they are mainly because developers may leak certain sensitive tokens to front-ends, which can be utilized by attackers to fake the identities (\textbf{T7}, \textbf{T8}). Hence, the developers have to be properly educated, and mandatory detection of such leakage should be enforced.  \looseness=-1

\paragraph{(L5) Vetting is not omnipotent} Mechanisms under vetted assumption  still need to be constantly supervised for malicious behaviors. We have identified 5 vetting-related mechanisms, where 4 are associated with threats. These threats involve both trust issues and vetting limitations. The major challenge super apps face is that malware actively acts against the vetting mechanism (\textbf{T12}), whereas vetting may not cover all potential vulnerabilities or malicious behaviors. Failure to realize so may result in trust issue, as super apps may trust vetted miniapps' behaviors, even though vetting may not reveal the capability for confusion or injection (\textbf{T4}, \textbf{T6}).  Even if a miniapp contains no malicious code during vetting, it still may be updated after being vetted if vulnerabilities allow them to hot-update code (\textbf{T5}). Hence, additional mechanisms to supervise the behavior of vetted miniapp need to be enforced.

\section{Trade-offs and Open Problems}
\label{sec:controversies}
% \begin{table}[]
%     \centering
%     \begin{tabular}{c|c}
%          &  \\
%          & 
%     \end{tabular}
%     \caption{Need a table illustrating the dilemmas and trade-offs}
%     \label{tab:dilemma}
% \end{table}
%\ZY{This can go to appendix, if the page limit is the concern}
\subsection{Trade-offs}
While third-party miniapps are allowed into super app ecosystems, it may still raise contradiction between the interest of super apps and users, especially when the super apps have collected a large amount of user data and open the access to third-party miniapps. While access control is enforced, it still may raise concern due to the \textit{``data sovereignty''}, where super apps may have sole rights to collect, control, dispose, and even delete the data and accounts of super app users. In this section, we discuss three trade-offs (\textbf{O}) derived from this dark side of super apps, on privacy issue accompanied by convenience, contradiction between user privacy and the platform's monetization utilizing privacy-sensitive data, and the debate between security and usability. In the end, we will summarize open problems (\textbf{P}) that may be addressed by future works based on our findings.

\paragraph{(O1) Privacy and Convenience} While users may experience convenience on seamless and convenient services such as food ordering by allowing miniapps to access personal information such as phone numbers to register accounts and place orders, risks occur when the privacy data of users are shared between miniapps and super apps. Nevertheless, the super app platforms need to seek a trade-off between user convenience and privacy protection, especially when these super apps involve data collected from millions of users.

\paragraph{(O2) Privacy and Monetization} Super apps may be tempted to monetize user data by sharing it with third parties for targeted advertising or other purposes. However, this poses privacy concerns, as users may not be aware of how their data is being used or whether they have control over data sharing. To mitigate these concerns, super apps should adopt transparent data practices, provide clear privacy policies, offer robust data protection measures, and give users control over their data, including the ability to opt out of data sharing or targeted advertising. 
% Compliance with applicable data protection regulations is also essential to ensure user privacy rights are respected. 
% However, these are only temporary solutions since it involves a compromise between user privacy and usability, which cannot be permanently resolved.

\paragraph{(O3) Security and Usability} 
Balancing security and usability is crucial in the super app ecosystem to optimize user experience while safeguarding user data and privacy. However, this task can be challenging as prioritizing security measures may hinder user experiences. For instance, users may opt for Bluetooth Low Energy (BLE) Secure Connections or Bluetooth Secure Simple Pairing (SSP) protocols to ensure secure authentication and pairing of devices. However, the process of pairing two devices together requires users to interact with the devices, such as entering passkeys and confirming the connection. Imposing these procedures can refrain users from using the super app altogether due to the added inconvenience.
% Although many super apps provide APIs (e.g., \texttt{wx.makeBluetoothPair} for {\sc WeChat}) to facilitate device pairing, miniapp developers may still choose not to utilize them, due to the concern of potentially negative impact on user satisfaction.

% \paragraph{(C5) User Awareness and Education} Users often rely heavily on super apps for various services, but they may not be fully aware of the security risks involved. Educating users about potential threats, safe usage practices, and privacy settings empowers them to make informed decisions and enhances overall security. For example,  the user may share sensitive information, such as his name, or credit card details, with unauthorized or untrusted miniapps. This makes it easier for attackers to gain unauthorized access or engage in identity theft. 

\subsection{Open Problems}
\paragraph{(P1) Security compliance analysis} In this paper, we have illustrated threats due to implementation issues and compatibility issues (\textbf{T1}, \textbf{T2}, \textbf{T3}, \textbf{T9}). As super apps may implement mechanisms differently, it is vital to analyze the security compliance, especially on evaluating whether the super apps offer protection mechanisms at least the same or higher level of security compared with underlying systems. Otherwise, the super app could become a favourable target for attackers to exploit users' devices and cloud-hosted data. Future works could include systematic analysis on existing platforms, and automatic tools for super app and developers to uncover potential vulnerabilities due to such discrepancy.   \looseness=-1

% Create tools to aid implementation of super apps
% and miniapps. Super app side, automatic compliance and
% verification tool to make sure compatibility of OSes. Devel-
% oper side, integrated check at testing side to reduce problems
% in vetting
% 12
\paragraph{(P2) Security mechanism standardization} We have revealed that the difference of implementation across super app platforms  may introduce vulnerabilities breaking access control (\textbf{T3}), or cause one platform to be  more vulnerable than the others (\textbf{T9}). Future works may include a study on measuring feasible security mechanisms for super app platforms, and a standardized security protection solution that can be deployed by all super app platforms to ensure that the protection implementations are equally stringent.

% together on a standardized security solution There are many implementation issues due to difference of implementation across platforms. All isolation related at super app side are due to implementation

\paragraph{(P3) Miniapp developer education} Besides the problems due to implementation of super apps, we also found that developers may leak essential but sensitive information to undesired parties (\textbf{T7}, \textbf{T8}). While the developers are not supposed to be all experts in miniapp security configuration, these developers need to be properly educated on recommended and prohibited practices. Therefore, super apps need to provide more developer-oriented highlight on security configuration in documentations. Future works may include automatic analysis and detection on leakage of privacy- and security-sensitive data that can be directly deployed at developers' ends or the vetting process to prevent vulnerable miniapps from being released to the ecosystem.

\paragraph{(P4) Semantic-aware miniapp vetting} In this paper, we discover that the current vetting mechanisms still face challenges and limitations despite the strict control on the distribution of miniapps (\textbf{T4}, \textbf{T5}, \textbf{T6}, \textbf{T10)}. Future work may be directed to modeling and formalizing super app specific miniapp malware, and automatic, semantic-aware detection techniques deployable by super apps to significantly thwart these malware from being distributed into the ecosystem.  \looseness=-1

\section{Related Work}
\label{sec:related}
% \paragraph{Miniapp Security}

% \paragraph{Extension Malware Detection}

\paragraph{Miniapp Security} The security of miniapps within super apps is currently receiving significant attention.
% Several studies have investigated different aspects of miniapp security. 
For instance, Lu \textit{et al.}~\cite{lu2020demystifying} uncovered resource management vulnerabilities that allow attackers to steal sensitive user information. Zhang \textit{et al.}~\cite{zhang2021measurement} developed MiniCrawler to analyze security practices, such as obfuscation usage and security-related API invocations of miniapps. More recently, Zhang \textit{et al.}~\cite{zhang2022identity}  discovered identity confusion vulnerabilities enabling attackers to exploit high-privileged capabilities. Additionally, Yang \textit{et al.}~\cite{cmrf} investigated vulnerabilities cross-miniapp channel in platforms like {\sc WeChat} and {\sc Baidu}. Wang \textit{et al.}~\cite{wang2022characterizing} proposed \textsc{WeDetector}, and identified 11 bugs from 25 real-world {\sc WeChat} miniapps. Wang \textit{et al.}~\cite{taintmini2023} introduce \textsc{TaintMini}, a comprehensive taint analysis framework to detect collusion attacks among miniapps. 
Zhang \textit{et al.}~\cite{appsecretleak} investigated the misuse and consequences of cryptographic keys, the ``AppSecret'', within miniapps. 
% , and explored the potential consequences of such attacks.
Wang \textit{et al.}~\cite{hiddenAPI2023} discovered previously undisclosed APIs offered by super apps and illustrated their potential for exploitation.  Wang \textit{et al.} ~\cite{crossplatform2023} systematically identified inconsistencies and derived vulnerabilities in {\sc WeChat} APIs across platforms.
% aiming to demonstrate their potential exploitation by remote attackers or malicious mini-apps. They developed \textsc{APIDiff}, a tool generating test cases and detecting execution disparities. They discovered three sets of discrepant APIs across diverse platforms, providing concrete examples of their exploitation. 
In contrast to prior efforts, our work represents the first systematic endeavor to systematize the security mechanisms, threats, and root cause of super apps.
% Additionally, we delve into the discussion of potential threats and their underlying causes. We propose open problems that require resolution and present valuable lessons learned from our study. 

\paragraph{App Ecosystem Security} In recent years, alongside mini-apps, various techniques have emerged to facilitate the development of a streamlined app ecosystem. For instance, Google Instant Apps, as mentioned by Aonzo \textit{et al.}~\cite{aonzo2018phishing}, have been introduced to support lightweight applications. Notably, Aonzo \textit{et al.}~\cite{aonzo2018phishing}   and Tang \textit{et al.} ~\cite{tang2020all} highlight the vulnerability of Google Instant Apps to be utilized in password-stealing attacks.   Super apps are evolved from web browsers. % do not come as a surprise, and they indeed developed from the web browsers. As such, prior efforts, particular for those focused on studying the components that used 
There is a large body of research focusing on browser security such as web extensions~\cite{xing2015understanding,some2019empoweb,chen2018mystique,wang2013unauthorized}, which are closely related to our study. For example, SABRE~\cite{dhawan2009analyzing} uses in-browser information-flow tracking to detects malicious browser extensions that leaks sensitive information. 
Hulk~\cite{kapravelos2014hulk} dynamically  detects malicious browser extensions by monitoring their execution.  Expector~\cite{xing2015understanding} inspects and identifies browser extensions that involve the advertisements, and then detects the malicious ones.  
%Different from those works, our study is the first to uncover the discrepancies in the super apps that can be exploited to mount attacks. 
\looseness=-1
%In contrast to these studies, our work introduces \sys, a comprehensive taint analysis framework specifically designed to detect collusion attacks among miniapps.

% \section{Availability}
% \label{sec:availability}
% \input{paper/09_availability.tex}

\section{Conclusion}
\label{sec:conclusion}

In conclusion, mobile supper apps have revolutionized the mobile app landscape by enabling third-party developers to deploy add-ons within these platforms. This study highlights the security measures implemented by super apps, including permissions, sandboxing, and encryption, to ensure data protection and system integrity. However, challenges such as phishing attacks and balancing revenue generation with user experience remain. User awareness and ongoing research are crucial for addressing these challenges and advancing the super app paradigm. Overall, the super app paradigm offers enhanced functionality and improved user experiences, but further efforts are needed to refine security models and mitigate potential threats.

\bibliographystyle{abbrv}
\bibliography{paper}

\appendix
% \usepackage{multirow}
 
\section{Super App Platforms Investigated}
\label{sec:yeye}
%\o/\o/\o/\o/\o/超哥\o/\o/\o/\o/

%^nihaonihao xinghuixinghui
In this paper, we collected 30 super app platforms with the keyword super app as listed in \autoref{tab:list}. These super apps are developed by companies based globally, including those in Asia (China, South Korea, Japan, India, Vietnam, Indonesia, Philippines, Singapore, and Saudi Arabia), Africa (South Africa and Kenya), Europe (Russia), and Latin America (Argentina). Through reverse engineering, we confirmed 20 platforms to be super app platforms with the integration of miniapps from third-party as marked in the blue rows of \autoref{tab:list}, where 4 super apps are implemented with Nebula framework from Alibaba, 1 super app with T7 core from Baidu, 3 super apps with Lynx from Bytedance, 3 super apps with X5 core from Tencent, and 3 super apps with Android WebView. While we have found similar rendering environments in apps claimed to be super apps, we did not find the entrance to miniapp-alike interfaces or the access is restricted due to regional issue (e.g., \texttt{Momo} only allows registration from Vietnamese phone number). Hence, we did not include these platforms as platforms studied in our paper. Interestingly, we found that some super apps such as \textsc{Yandex} have created Github repositories related to miniapp demos~\cite{yandexfront, yandexback}, which may indicate that the miniapp frameworks are under development and testing. Hence, more super apps may emerge after these platforms have finally released the miniapp framework.

\begin{table*}[]
 \scriptsize
\centering
\setlength{\tabcolsep}{8pt}
\begin{tabular}{l|l|r|c|c|l}
\toprule
          \textbf{Platform}& \textbf{Country}      & \textbf{Users}\textbf{ (M)} & \textbf{Rendering framework}         & \textbf{Has miniapp?} & Note                       \\ \hline
\rowcolor[HTML]{CFE2F3}AliPay~\cite{alipayuser}       & China        & 1,200        &     \multirow{4}{*}{Nebula}                             & \tickYes             &                            \\
\rowcolor[HTML]{CFE2F3}DingTalk~\cite{dingtalk}     & China        & 600         &      Nebula     & \tickYes             &                            \\
\rowcolor[HTML]{CFE2F3}MPesa~\cite{mpesauser}        & Kenya        & 52          &                                                     & \tickYes             &                            \\
\rowcolor[HTML]{CFE2F3}VodaPay~\cite{vodapayuser}      & South Africa & 3         &                                                     & \tickYes             &                            \\\hline
\rowcolor[HTML]{CFE2F3}Baidu~\cite{baiduuser}        & China        & 657         & T7 core                                                   & \tickYes             &                            \\\hline
% ByteDance    & China        &              & \multirow{4}{*}{Lynx}                                 & \tickYes             &                            \\
\rowcolor[HTML]{CFE2F3}Huoshan      & China        &       -       &      \multirow{3}{*}{Lynx}                             & \tickYes             &                            \\
\rowcolor[HTML]{CFE2F3}Tiktok~\cite{tiktokuser}       & China        & 1,700        &   Lynx                               & \tickYes             &                            \\
\rowcolor[HTML]{CFE2F3}Toutiao~\cite{toutiaouser}      & China        & 209         &             & \tickYes             &                            \\\hline
\rowcolor[HTML]{CFE2F3}QQ~\cite{qquser}           & China        & 597         &  \multirow{3}{*}{XWeb w/ X5 core}                                & \tickYes             &                            \\
\rowcolor[HTML]{CFE2F3}WeChat~\cite{wechatuser}       & China        & 1,300        &            XWeb w/ X5 core                      & \tickYes             &                            \\
\rowcolor[HTML]{CFE2F3}WeCom~\cite{wecomuser}        & China        & 73         &  & \tickYes             &                            \\ \hline
\rowcolor[HTML]{CFE2F3}Kuaishou~\cite{kuaishouuser}     & China        & 347         &                           Android WebView                                & \tickYes             &                            \\\hline
Meituan~\cite{meituanuser}      & China       & 680         &                         React Native                              &               &                            \\\hline

Momo~\cite{momouser}         & Vietnam      & 115         & \multirow{2}{*}{React Native}                                                &               & Registration restricted    \\
Ola~\cite{olauser}          & India        & 200         &                                               &               & Miniapp entrance not found\\ \hline
Paypay~\cite{paypayuser}       & Japan        & 55          & \multirow{13}{*}{Android WebView}            
&               & Registration restricted    \\
\rowcolor[HTML]{CFE2F3}Line~\cite{lineuser}        & South Korea  & 230         &                                            & \tickYes             &                            \\
Toss~\cite{tossuser}         & South Korea  & 21          &                                            &               & Miniapp entrance not found \\
Kakao~\cite{kakaouser}        & South Korea  & 53          &                                            &               & Registration restricted    \\
\rowcolor[HTML]{CFE2F3}Zalo~\cite{zalouser}         & Vietnam      & 75          &                                            & \tickYes             &                            \\

Paytm~\cite{paytmuser}        & India        & 330         &                                            &               & Anti-analysis enforced     \\
Airtel Money~\cite{airteluser} & India        & 10          &                                            &               & Miniapp entrance not found \\
% Miniapp entrance not found \\
Gojek        & Indonesia    & 38          &                                            &               & Miniapp entrance not found \\
PayMaya~\cite{paymayauser}      & Philipine    & 44          &                                            &               & Miniapp entrance not found \\
Grab~\cite{grabuser}         & Singapore    & 33          &                                            &               & Miniapp entrance not found \\
Careem~\cite{careemuser}       & Saudi Arabia & 48          &                                            &               & Miniapp entrance not found \\
Jumia~\cite{jumiauser}        & Nigeria      & 2         &                                            &               & Miniapp entrance not found \\
Mercado Pago~\cite{mercadouser} & Argentina    & 20          &                                            &               & Miniapp entrance not found \\\hline
Yandex~\cite{yandexuser}       & Russia       & 55          & Chromium                                                  &               & Miniapp entrance not found \\\hline

\rowcolor[HTML]{CFE2F3}Binance~\cite{binanceuser}      & China        & 128             &                          Android WebView                                 & \tickYes             &                                     \\
\bottomrule
\end{tabular}

\caption{Summary of platforms investigated in this paper.}
\label{tab:list}
\end{table*}
 
\end{document}